\documentclass[12pt, draftclsnofoot, onecolumn]{IEEEtran}
\newcommand\figurescaling{0.45}
\usepackage[utf8]{inputenc}
\usepackage[english]{babel}
\usepackage[T1]{fontenc}

\usepackage{amsmath}
\usepackage{amsfonts}
\usepackage{mathtools}
\usepackage{enumerate}
\usepackage{booktabs}
\usepackage[table]{xcolor}
\usepackage{multicol}
\usepackage{multirow}
\usepackage{graphicx}
\usepackage[shortlabels]{enumitem}
\usepackage{soul}
\usepackage[acronym]{glossaries}
\usepackage{pgfplots}
\pgfplotsset{compat=1.15}
\usepackage{tikz}
\usetikzlibrary{patterns}
\usepackage[outline]{contour}
\contourlength{1.8pt}
\usepackage{caption}
\usepackage{subcaption}
\usepackage{floatrow}%

\usepackage[acronym]{glossaries}
\newacronym{phy}{PHY}{Physical}
\newacronym{mac}{MAC}{Medium Access Control}
\newacronym{ns}{NS}{Network Server}
\newacronym{gw}{GW}{Gateway}
\newacronym{ed}{ED}{End Device}
\newacronym{adr}{ADR}{Adaptive Data Rate}
\newacronym{sf}{SF}{Spreading Factor}
\newacronym{cr}{CR}{Code Rate}
\newacronym{mcs}{MCS}{Modulation and Coding Scheme}
\newacronym{ack}{ACK}{Acknowledgment}
\newacronym{iot}{IoT}{Internet of Things}
\newacronym[plural=LPWANs,firstplural=Low Power Wide Area Networks (LPWANs)]{lpwan}{LPWAN}{Low Power Wide Area Network}
\newacronym{uplink}{UL}{uplink}
\newacronym{toa}{ToA}{time on air}
\newacronym{downlink}{DL}{downlink}
\newacronym{qos}{QoS}{Quality of Service}
\newacronym{css}{CSS}{Chirp Spread Spectrum}
\newacronym{dc}{DC}{Duty Cycle}
\newacronym{rx1}{RX1}{first receive window}
\newacronym{rx2}{RX2}{second receive window}
\newacronym{sb1}{SB1}{Sub-Band 1}
\newacronym{sb2}{SB2}{Sub-Band 2}
\newacronym{fdgw}{FDGW}{Full Duplex Gateway}
\newacronym{ttn}{TTN}{The Things Network}
\newacronym{ul}{UL}{Uplink}
\newacronym{dl}{DL}{Downlink}
\newacronym{ism}{ISM}{Industrial, Scientific, and Medical}
\newacronym{cpsr}{CPSR}{Confirmed Packet Success Rate}
\newacronym{der}{DER}{Data Extraction Rate}
\newacronym{cder}{CDER}{Confirmed Data Extraction Rate}
\newacronym{uder}{UDER}{Unconfirmed Data Extraction Rate}
\newacronym{uu}{UU}{Unconfirmed Uplink PDR}
\newacronym{cu}{CU}{Confirmed Uplink PDR}
\newacronym{cd}{CD}{Confirmed Downlink PDR}
\newacronym{pdr}{PDR}{Packet Delivery Rate}

\title{A Configurable Mathematical Model for Single-Gateway LoRaWAN
  Performance Analysis} \author{Davide Magrin, \textit{Member,
    IEEE},
  Martina Capuzzo, \textit{Student Member, IEEE}, \\
  Andrea Zanella, \textit{Senior Member, IEEE}, and Michele Zorzi,
  \textit{Fellow, IEEE} \thanks{The authors are with the Dept.\ of
    Information Engineering (DEI), University of Padova, Via Gradenigo 6/b,
    35131 Padova, Italy. Profs. A. Zanella and M. Zorzi are also with the
    Human Inspired Technology (HIT) center of the University of Padova and
    with the Italian National Inter-University Consortium for
    Telecommunication (CNIT). Email: \{magrinda, capuzzom, zanella,
    zorzi\}@dei.unipd.it}}

\usepackage{fancyhdr}
\fancypagestyle{FirstPage}{
  \chead{This work has been submitted to the IEEE for possible publication.
    Copyright may be transferred without notice,
    after which this version may no longer be accessible.}
}

\begin{document}

\maketitle

\thispagestyle{FirstPage}

\begin{abstract}

  LoRaWAN is a Low Power Wide Area Network technology featuring long
  transmission ranges and a simple MAC layer, which can support sensor data
  collection, control applications and reliable services thanks to the
  flexibility offered by a large set of configurable system parameters. However,
  the impact of such parameters settings on the system's performance is often
  difficult to predict, depending on several factors. To ease this task, in this
  paper, we provide a mathematical model to estimate the performance of a
  LoRaWAN gateway serving a set of devices that may or may not employ confirmed
  traffic. The model features a set of parameters that can be adjusted to
  investigate different gateway and end-device configurations, making it
  possible to carry out a systematic analysis of various trade-offs. The results
  given by the proposed model are validated through realistic ns-3 simulations
  that confirm the ability of the model to predict the system performance with
  high accuracy, and assess the impact of the assumptions made in the model for
  tractability.

\end{abstract}


\section{Introduction}
\label{sec:introduction}

The \gls{iot} makes it possible to remotely monitor and control a wide set of
heterogeneous objects through an Internet connection. This paradigm foresees
multiple applications in a large variety of scenarios: from fleet tracking and
process monitoring in industrial scenarios to smarter garbage collection and
intelligent light control in cities; from monitoring of soil moisture in
agriculture to home temperature control and personal health monitoring
\cite{zanella2014internet, yuehong2016internet, dlodlo2015internet,
  chiariotti2018symbiocity}. Also, the \gls{iot} paradigm can be applied to
surveillance-related applications~\cite{bovenzi2018iot}, as event detectors and
alarms~\cite{dos2020performance}.

The presence of several use cases spawned an ample market, and encouraged the
development of multiple technologies meeting the need for low-cost ubiquitous
connectivity. A large part of \gls{iot} nodes will consist in sensors that
generate sporadic traffic, without strict constraints in terms of latency and
throughput. This calls for new wireless solutions able to support a massive
number of devices, with an affordable cost for both user equipment and network
infrastructure. Therefore, high energy efficiency, extended coverage, and
infrastructure simplicity are aspects of primary importance.

Such requirements motivated the creation of a new family of wireless
technologies collectively called \glspl{lpwan}, characterized by long
coverage range and low power consumption. A prominent \gls{lpwan}
technology is LoRaWAN, which claims up to 10~years of battery lifetime for
devices, and a transmission range between 1.5~km in urban scenarios and
30~km in rural areas~\cite{centenaro2016long}.



Since the deployment of a dense \gls{iot} network is expensive and time
consuming, performance assessments using simulations and mathematical models
become essential to gauge the effect of network parameters and estimate the
performance at a reduced cost. In this work, we propose an analytical model of
the performance of a LoRaWAN network, accounting for the most relevant features
of the LoRaWAN standard. This model considers network-layer performance,
assuming perfect orthogonality between signals modulated with different
\glspl{sf}. However, compared to previous models in the literature (discussed in
Sec. III), our model takes into account a wider range of aspects of the PHY and
MAC layers, such as the possibility of transmitting multiple times both
confirmed and unconfirmed packets, the limitations on the channel occupancy time
imposed by the different national regulations, the interference produced by
multiple overlapping transmissions, the capture effect, and the limited number
of demodulators available at the \gls{gw}. Furthermore, the model formulation
offers great flexibility in setting some system parameters, thus making it
possible to analyze the system performance under different conditions and to
shed light on possible trade-offs. We consider as performance
metrics the packet success probability, average delays, and fairness, from which
it is possible to derive other measures of interest, such as energy consumption,
system's reliability and the achievable \gls{qos} in multiple scenario.
%
The proposed model is validated by comparing the results with those obtained
through detailed ns-3 simulations. The analysis shows how the model can be used
to maximize different performance metrics, proving a very powerful and
convenient tool to determine the best network configuration.


The rest of this work is structured as follows. To make the paper
self-contained, in Sec.~\ref{sec:technology} we present the main features of the
LoRaWAN standard, while in Sec.~\ref{sec:soa} we give an overview of the current
state of the art in the performance modeling of this technology.
Sec.~\ref{sec:model} introduces the proposed model and describes how some of its
parameters can be tuned to explore different behaviors of the network, while
Sec.~\ref{sec:simulation} briefly describes the simulation framework used for
validation. Sec.~\ref{sec:results}, then, compares the output of the analytical
and simulation models, also showing how they can be used to provide different
insights of the network behavior. Finally, Sec.~\ref{sec:conclusion} draws the
conclusions and discusses possible future developments.


\section{Technology overview}
\label{sec:technology}

This section describes the key LoRaWAN features, dwelling upon the elements and
properties that have a significant impact on the system-level performance, which
will then be considered in the model formulation.

\subsection{The LoRa modulation}
\glsreset{sf}

LoRa is a modulation technique based on \gls{css}, patented by Semtech. Bitrate
and coverage range depend on the \gls{sf} parameter that can vary from 7 to 12. Lower \gls{sf}
values achieve higher data rates and shorter transmission times, but require
higher signal powers at the receiver for correct decoding, which implies shorter
coverage ranges. On the other hand, signals transmitted using higher \gls{sf}
values are more robust to channel impairments and can thus achieve longer
transmission distances, at the price of an increased transmission time due to
their lower data rates. Furthermore, signals modulated with different \glspl{sf}
are almost orthogonal: even if overlapping in time and frequency, two or more
signals transmitted with different \glspl{sf} can be simultaneously decoded,
provided that their received powers satisfy some
conditions~\cite{croce2018impact}.

When multiple packets transmitted with the same \gls{sf} overlap in time and
frequency, instead, they may generate destructive mutual interference,
disrupting each other's reception and resulting in what is called a
\textit{packets collision} event. However, if one signal is significantly
stronger than the others, by a power margin greater than the so-called
``co-channel rejection parameter'' $CR_{\rm dB}$, then it can be received
correctly despite the interference, giving rise to a \emph{capture} phenomenon.

The value of $CR_{dB}$ has been estimated to be around 6~dB
in~\cite{goursaud2015dedicated}. In order to take advantage of these features,
the SX1301 LoRa PHY chipset, typically employed in \glspl{gw}~\cite{sx1301},
provides 8 parallel demodulation chains, which allow the chip to demodulate up
to 8 different signals simultaneously, irrespective of their \glspl{sf} and
frequency. We also remark that the \glspl{gw} do not support full-duplex
transmission and reception: in order to send a \gls{dl} packet they have to
interrupt any ongoing reception, regardless of the frequency channels in which
transmission and reception occur.


\subsection{The LoRaWAN standard}

\begin{figure}[t]
  \centering
  \includegraphics[width=\figurescaling\linewidth]{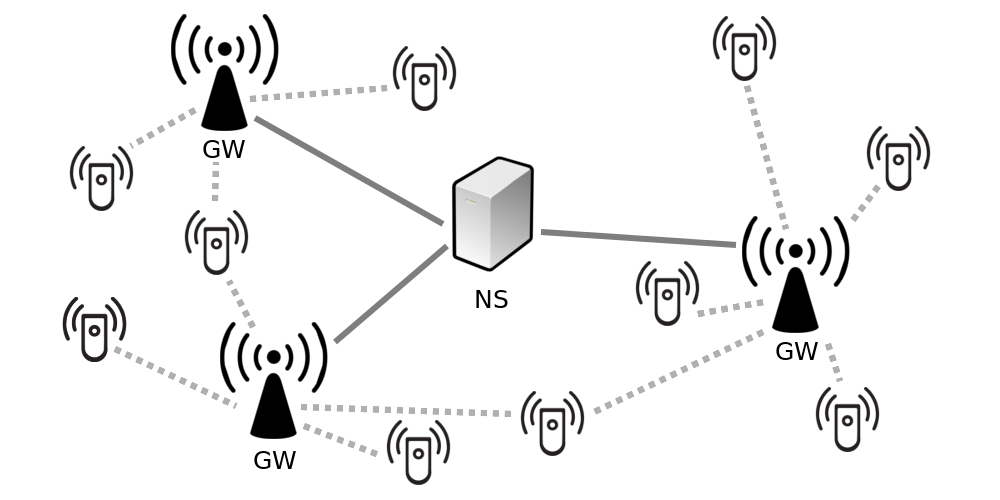}
  \caption{LoRaWAN network infrastructure. Dotted lines represent LoRa links,
    while solid lines represent IP connections.}
  \label{fig:infrastruc}
  \vspace{-1em}
\end{figure}

\glsreset{ns} \glsreset{ed} \glsreset{gw} The LoRaWAN standard~\cite{lorawan}
defines a star-of-stars topology, as represented in Fig.~\ref{fig:infrastruc},
with three kinds of devices: the \textit{\gls{ns}}, which is the central network
controller and can be located anywhere in the Internet; the \textit{\glspl{ed}},
peripheral nodes (usually sensors or actuators) that transmit using the LoRa
modulation; and the \textit{\glspl{gw}}, relay nodes that collect messages from
the \glspl{ed} through the LoRa interface and forward them to the \gls{ns} using
a reliable IP connection, and \textit{vice versa}.

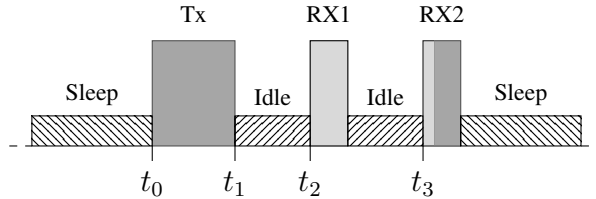
\begin{figure}[t]
  \centering


\begin{tikzpicture}

  \def\bottomy{0};
  \def\dashedsidelength{0.3};
  \def\blockheightshort{0.4};
  \def\blockheighttall{1.4};
  \def\blocksleepwidth{1.6};
  \def\blocktxwidth{1.1};
  \def\blockstandbywidth{0.};
  \def\blockidlewidth{1.};
  \def\blockrxwinwidth{0.5};
  
  \def\ybottom{1.};
  \def\textbottom{0};

  \draw [dashed] (0,\ybottom) -- (\dashedsidelength,\ybottom);
  \draw [pattern=north west lines](\dashedsidelength,\ybottom) rectangle node[above, yshift=0.2cm]{\footnotesize{Sleep}} ({\blocksleepwidth + \dashedsidelength},{\ybottom + \blockheightshort});
  
  \draw [fill=gray, opacity=0.7](\dashedsidelength+\blocksleepwidth,\ybottom) rectangle node[text=black, opacity=1, above, yshift=0.8cm]{\footnotesize{Tx}} ({\blocktxwidth + \blocksleepwidth + \dashedsidelength},{\ybottom + \blockheighttall});
  \draw (\dashedsidelength+\blocksleepwidth,\ybottom) -- (\dashedsidelength+\blocksleepwidth,\ybottom-0.2) node {};
  \node[yshift=-0.5cm] at (\dashedsidelength + \blocksleepwidth, \ybottom) {$t_{0}$};
  \draw (\blocktxwidth + \dashedsidelength + \blocksleepwidth, \ybottom) -- (\blocktxwidth + \dashedsidelength + \blocksleepwidth, \ybottom-0.2) node {};
  \node[yshift=-0.5cm] at (\blocktxwidth + \dashedsidelength + \blocksleepwidth, \ybottom) {$t_{1}$};

  \draw (\dashedsidelength+\blocksleepwidth + \blocktxwidth,\ybottom) rectangle node[above, yshift=0.2cm]{\footnotesize{}} ({\blockstandbywidth + \blocktxwidth + \blocksleepwidth + \dashedsidelength},{\ybottom + \blockheightshort});
  \draw [fill=gray, opacity= 0.3](\dashedsidelength+\blocksleepwidth + \blocktxwidth,\ybottom) rectangle node[above, yshift=0.2cm]{\footnotesize{}} ({\blockstandbywidth + \blocktxwidth + \blocksleepwidth + \dashedsidelength},{\ybottom + \blockheightshort});
  
  \draw [pattern = north east lines](\blockstandbywidth + \dashedsidelength+\blocksleepwidth + \blocktxwidth,\ybottom) rectangle node[above, yshift=0.2cm]{\footnotesize{Idle}} ({\blockidlewidth + \blocktxwidth + \blocksleepwidth + \dashedsidelength},{\ybottom + \blockheightshort});
  
  \draw (\blockidlewidth + \dashedsidelength+\blocksleepwidth + \blocktxwidth,\ybottom) rectangle node[text=black, opacity=1, above, yshift=0.8cm]{\footnotesize{RX1}} ({\blockrxwinwidth + \blockidlewidth + \blocktxwidth + \blocksleepwidth + \dashedsidelength},{\ybottom + \blockheighttall});
  \draw (\blockidlewidth + \dashedsidelength+\blocksleepwidth + \blocktxwidth,\ybottom) -- (\blockidlewidth + \dashedsidelength+\blocksleepwidth + \blocktxwidth,\ybottom-0.2) node {};
  \node[yshift=-0.5cm] at (\blockidlewidth + \dashedsidelength+\blocksleepwidth + \blocktxwidth,\ybottom) {$t_{2}$};
  \draw [fill=gray, opacity= 0.3, text=black](\blockidlewidth + \dashedsidelength+\blocksleepwidth + \blocktxwidth,\ybottom) rectangle node[text=black, opacity=1, above, yshift=0.8cm]{\footnotesize{}} ({\blockrxwinwidth + \blockidlewidth + \blocktxwidth + \blocksleepwidth + \dashedsidelength},{\ybottom + \blockheighttall});
  
  \draw [pattern = north east lines](\blockrxwinwidth + \blockidlewidth + \dashedsidelength+\blocksleepwidth + \blocktxwidth,\ybottom) rectangle node[text=black, above, yshift=0.2cm]{\footnotesize{Idle}} ({\blockrxwinwidth+2*\blockidlewidth + \blocktxwidth + \blocksleepwidth + \dashedsidelength},{\ybottom + \blockheightshort});

  \draw (\blockrxwinwidth + 2*\blockidlewidth + \dashedsidelength+\blocksleepwidth + \blocktxwidth,\ybottom) rectangle node[text=black, opacity=1, above, yshift=0.8cm]{\footnotesize{RX2}} ({2*\blockrxwinwidth + 2*\blockidlewidth + \blocktxwidth + \blocksleepwidth + \dashedsidelength},{\ybottom + \blockheighttall});
  \draw (\blockrxwinwidth + 2*\blockidlewidth + \dashedsidelength+\blocksleepwidth + \blocktxwidth,\ybottom) -- (\blockrxwinwidth + 2*\blockidlewidth + \dashedsidelength+\blocksleepwidth + \blocktxwidth,\ybottom-0.2) node {};
  \node[yshift=-0.5cm] at (\blockrxwinwidth + 2*\blockidlewidth + \dashedsidelength+\blocksleepwidth + \blocktxwidth,\ybottom) {$t_{3}$};
  \draw [draw=none, fill=gray, opacity= 0.3](\blockrxwinwidth + 2*\blockidlewidth + \dashedsidelength+\blocksleepwidth + \blocktxwidth,\ybottom) rectangle node[above, yshift=0.8cm]{\footnotesize{}} ({1.3*\blockrxwinwidth + 2*\blockidlewidth + \blocktxwidth + \blocksleepwidth + \dashedsidelength},{\ybottom + \blockheighttall});
  \draw [draw=none, fill=gray, opacity= 0.7](1.3*\blockrxwinwidth + 2*\blockidlewidth + \dashedsidelength+\blocksleepwidth + \blocktxwidth,\ybottom) rectangle node[above, yshift=0.8cm]{\footnotesize{}} ({2*\blockrxwinwidth + 2*\blockidlewidth + \blocktxwidth + \blocksleepwidth + \dashedsidelength},{\ybottom + \blockheighttall});

  \draw (2*\blockrxwinwidth + 2*\blockidlewidth + \dashedsidelength+\blocksleepwidth + \blocktxwidth,\ybottom) rectangle node[text=black, above, yshift=0.2cm]{\footnotesize{}} ({\blockstandbywidth + 2*\blockrxwinwidth + 2*\blockidlewidth + \blocktxwidth + \blocksleepwidth + \dashedsidelength},{\ybottom + \blockheightshort});
  \draw [fill=gray, opacity= 0.3](2*\blockrxwinwidth + 2*\blockidlewidth + \dashedsidelength+\blocksleepwidth + \blocktxwidth,\ybottom) rectangle node[text=black, above, yshift=0.2cm]{\footnotesize{}} ({\blockstandbywidth + 2*\blockrxwinwidth + 2*\blockidlewidth + \blocktxwidth + \blocksleepwidth + \dashedsidelength},{\ybottom + \blockheightshort});

  \draw [pattern=north west lines](2*\blockrxwinwidth + 2*\blockidlewidth + \dashedsidelength+\blocksleepwidth + \blocktxwidth + \blockstandbywidth,\ybottom) rectangle node[above, yshift=0.2cm]{\footnotesize{Sleep}} ({\blockstandbywidth + 2*\blockrxwinwidth + 2*\blockidlewidth + \blocktxwidth + 2*\blocksleepwidth + \dashedsidelength},{\ybottom + \blockheightshort});
  
  \draw [dashed] ({\blockstandbywidth + 2*\blockrxwinwidth + 2*\blockidlewidth + \blocktxwidth + 2*\blocksleepwidth + \dashedsidelength},\ybottom) -- ({\blockstandbywidth + 2*\blockrxwinwidth + 2*\blockidlewidth + \blocktxwidth + 2*\blocksleepwidth + 2*\dashedsidelength},\ybottom);

\end{tikzpicture}
      \caption{Example of operational phases for a Class-A \gls{ed}. The device,
        initially in sleep mode, wakes up to transmit from time
        $t_0$ to $t_1$. Then, the node remains in the idle state for 1 second, and
        at time $t_2=t_1+1~s$ opens the \acrfull{rx1}. If no packets are
        received, the device remains idle until the \acrfull{rx2} is opened at
        time $t_3=t_1+2~s$. }
  \label{fig:cycle}
  \vspace{-1em}
\end{figure}

The standard also defines three classes of \glspl{ed}, which differ for the time
they spend in reception mode. This article considers \textit{Class A} devices,
which have the strictest requirements in terms of energy consumption. In order
to save battery, these devices stay in sleep mode most of the time, opening two
reception windows only 1 and 2 seconds after the end of an \gls{ul} packet
transmission. Fig.~\ref{fig:cycle} depicts the different operational phases of a
Class-A device when transmitting a \gls{ul} packet and, then, receiving a
\gls{dl} packet in the second receive window.

The \glspl{ed} have the possibility of transmitting \textit{unconfirmed} or
\textit{confirmed} packets. In the latter case, an \gls{ack} is expected in one
of the two reception opportunities after the transmission to confirm the correct
reception of the packet by the \gls{ns}.\footnote{Although in this paper we
  focus on ACK transmissions, the model and the analysis equally hold for any
  \gls{dl} packet returned by the \gls{ns} to the ED after the reception of a
  \gls{ul} packet by the NS.} If the \gls{ack} is not received in either of the
two reception windows, a re-transmission can be performed at least
RETRANSMIT\_TIMEOUT seconds after the end of the second receive window. The
standard recommends to randomly pick the value for RETRANSMIT\_TIMEOUT
uniformly between 1 and 3 seconds~\cite{regional}.
If the \gls{ack} is not received, the same confirmed message can be
re-transmitted up to $m-1$ times, after which the packet is dropped. The value
of $m$ can be configured by the \gls{ns}.\footnote{This behavior holds for the
  LoRaWAN 1.1 standard~\cite{lorawan, regional}: other versions of the standard
  may differ.} To increase the robustness of unconfirmed transmissions, instead,
the \gls{ed} can transmit each packet $h$ times. Once again, the value of $h$
can be set by the \gls{ns}. It is worth noting that the reception windows are
opened after every UL transmission, irrespective of whether or not an \gls{ack}
is expected, in order to give the \gls{ns} the opportunity to send a \gls{dl}
packet to the \glspl{ed}, if needed. The \gls{ul} messages transmitted by an
\gls{ed} are collected by all the \glspl{gw} in the coverage range of the
transmitter, and forwarded to the \gls{ns}. If the \gls{ed} requires a reply,
the \gls{ns} can pick any of these \glspl{gw} to transmit the \gls{dl} message.

LoRaWAN operates in the ISM unlicensed spectrum, the use of which is subject to
national regulations that define the maximum transmit power, and
the \gls{dc} limit, i.e., the maximum percentage of time a node can actively
transmit on a certain frequency band. The frequency bands, power and \gls{dc}
restrictions that apply to different regions are reported in the
standard~\cite{regional}. In particular, Table~\ref{tab:channels}
shows the configuration mandated for the European region, which entails three
bidirectional channels and a fourth channel reserved to \gls{dl} transmissions
only. The 868.1, 868.3, 868.5~MHz channels belong to the same regulatory
sub-band (\gls{sb1}), and have to share a \gls{dc} limitation of 1\%, while the
channel reserved for \gls{dl}, located in the 869~MHz sub-band (\gls{sb2}), can
benefit from a more lenient \gls{dc} of 10\% and a higher transmission power.

\begin{table}[h]
  \footnotesize
  \centering
  \caption{Available LoRaWAN channels in the two sub-bands.}
  \label{tab:channels}
  \begin{tabular}{llrr}
    \toprule
    Sub-band & Frequency [MHz] & Use & Duty Cycle \\
    \midrule
    \multirow{3}{*}{SB1} & 868.1 & UL/DL & 1\%, shared \\
    & 868.3 & UL/DL & 1\%, shared \\
    & 868.5 & UL/DL & 1\%, shared \\
    \arrayrulecolor{black!70}\midrule
    SB2 & 869.525 & DL & 10\%, dedicated \\
    \arrayrulecolor{black}\bottomrule
  \end{tabular}
\end{table}

The \gls{sf} used for a device’s transmission is configured by the device itself
or set by the \gls{ns} according to some network management policies. By
default, \glspl{ed} open the \gls{rx1} on the same frequency channel of the
\gls{ul} transmission, and expect a signal modulated with the same \gls{sf}. The
\gls{rx2}, instead, is opened on the 869.525~MHz channel and the incoming signal is
assumed to use \gls{sf} 12, to maximize the coverage rate. The standard allows
the \gls{ns} to modify this pre-defined configuration by communicating the new
settings to the \gls{ed} through appropriate \gls{mac} commands, allowing for
the use of any \gls{sf} in the second window.

\subsection{Packet life cycle}
\label{sec:lifecycle}

Messages transmitted by \glspl{ed} to the \gls{gw} are subject to multiple
causes of losses:

\begin{itemize}
\item \textit{Interference}: packets sent in the same frequency channel and
  with the same \gls{sf} collide. A transmission can survive a collision
  event if its received power is sufficiently higher than that of the other
  overlapping signals (capture effect).
\item \textit{\gls{gw} already in transmission}: the \gls{gw} can not lock
  on a \gls{ul} packet while performing a \gls{dl} transmission.
\item \textit{\gls{gw} starting a transmission}: an ongoing packet
  reception may be interrupted if the \gls{gw} needs to send a \gls{dl}
  packet.
\item \textit{No available demodulation chains at the \gls{gw}}: all demodulators
  are already busy decoding incoming signals.
\end{itemize}

Moreover, confirmed \gls{ul} messages cause the \gls{ns} to generate
\glspl{ack} that need to be transmitted by the \gls{gw}. Such DL transmissions
may as well be impaired by a number of events:
\begin{itemize}
\item \textit{Unavailability of receive windows}: this event occurs when
  all available \glspl{gw} are prevented from transmitting in both the
  receive windows because of the \gls{dc} constraint or other ongoing
  transmissions.
\item \textit{Interference}: \gls{dl} packets transmitted in \gls{rx1} can
  collide with \gls{ul} packets transmitted by other \glspl{ed} in the same
  channel and with the same \gls{sf}.
\end{itemize}

In this work, we provide a network model that accounts for all these
events.


\section{State of the art in LoRaWAN modeling}
\label{sec:soa}

In the last years, 
mathematical modeling has been applied to assess the network
performance with respect to various metrics. In~\cite{adelantado2017under}, the
authors address high-level questions about LoRaWAN's suitability for a range of
smart city applications, from metering to video surveillance, by modeling the
system as a superposition of different Aloha networks. They conclude that, even if
the long coverage range of a single \gls{gw} makes the infrastructure able to
serve several devices, the network must be carefully dimensioned to meet the
application requirements. The work presented in~\cite{georgiou2017low} is one of
the first to address the issue of scalability, using stochastic geometry to
model interference in a LoRaWAN network. However, the study considers scenarios
with only \gls{ul} traffic.
In~\cite{sorensen2017analy} instead, queueing theory is applied to model latency
and throughput of an \gls{ed} subject to \gls{dc} constraints, again focusing on
\gls{ul} communication only. The authors of~\cite{bankov2017mathem,
  bankov2019lorawan} provide a model based on Poisson arrival processes which
takes \gls{dl} communications, re-transmissions and capture effect into account.
However, the analysis holds only in limited-size networks, where nodes can
employ any transmission rate and their received powers are similar. The authors
of~\cite{croce2019lora} consider the features of the technology at the \gls{phy}
layer, by focusing on the capture effect and imperfect orthogonality between
\glspl{sf}: after performing empirical measurements, they model these effects
and derive the throughput achieved by the network for different cell
configurations and number of \glspl{gw}. In~\cite{heusse2020capacity}, the
problem of network scalability is faced through mathematical modeling and
Python-based simulations, taking into account also the capture effect, and
evaluating the impact of \gls{sf} allocation and power control. In all these
works, however, the main focus is on the \gls{phy} layer, and downlink traffic
and re-transmissions are not considered. Finally, the work presented
in~\cite{khan2019model} proposes a model to calculate energy consumption and
delay for reliable \gls{ul} traffic in a LoRaWAN network. The results for a
limited number of devices are compared to real test-bed measurements and to the
outcome of ns-3 simulations. The analysis, based on Markov-chain theory,
neglects the \gls{dc} constraints in the different sub-bands, and assumes that
\glspl{ack} are always sent in one specific receive window (either \gls{rx1} or
\gls{rx2}). Markov chains are also proposed in~\cite{delgado2021batteryless} to
characterize the performance of a LoRaWAN battery-less device; however, the
study considers a single device, and the network analysis is left for future
work.

The work presented in this paper is an extension of our previous conference
paper~\cite{capuzzo2018mathematical}, where we modeled a wide network with
Poisson packet arrivals, considering the \gls{dc} limitations and a set of
network parameters. Here we revise the model by developing a novel approach to
accurately consider the limited availability of reception chains at the \gls{gw},
the peculiarities of the two receive windows, and the \gls{dc} constraints.
Additionally, we include packet re-transmissions and the capture
effect. Compared to the state of the art, our model includes the ability to take
into account the coexistence of unconfirmed and confirmed traffic and, at the
same time, maintains the possibility of estimating the network behavior under
several network configurations with minimal effort. The results obtained through
this model are compared with those given by a state-of-the-art and open source
LoRaWAN simulator, presented in~\cite{magrin2017performance}, further attesting
the accuracy of the proposed approach and exploring the impact of common
assumptions. Finally, we also show some possible usages of the model to evaluate
a wide variety of network configurations with limited effort.


\section{Model}
\label{sec:model}

The aim of the model proposed in this paper is to characterize the behavior of a
LoRaWAN network with a single \gls{gw}, which receives packets from a set of
\glspl{ed} and needs to reply in one of the two receive windows when an \gls{ed}
requires confirmation. The system performance is assessed in terms of packet
success probability, following the approach used
in~\cite{capuzzo2018mathematical} and extending it with a more accurate
characterization of the \gls{gw} behavior. This performance metric is proxy to
other fundamental metrics, such as throughput and network capacity, which can be
straightforwardly derived from it. The following sub-sections are structured as
follows. The reference scenario, model assumptions, system parameters and their
effects are described in Sec.~\ref{sec:scenario}, together with a brief
presentation of the structure of the model and its underlying rationale;
Sec.~\ref{sec:quantities}, then, describes some relevant quantities and
parameters of the proposed model. We then delve into the analytical formulation
by decoupling the analysis of the \gls{ul} traffic
(Sec.~\ref{sec:ultrafficrates} and~\ref{sec:phy-probs}) and \gls{dl} messages
(Sec.~\ref{sec:acktransmission}), and derive the formulas for \gls{dl} success
probabilities in Sec.~\ref{sec:succprobs}. Finally, Sec.~\ref{sec:metrics},
describes different performance metrics and their computation. Note that,
because of the mutual dependency of some values, some terms may be described and
introduced before the corresponding equation can be derived, in which case
references are provided in the text.

\subsection{Scenario and assumptions}
\label{sec:scenario}

We consider a scenario where the \glspl{ed} are randomly and uniformly
distributed around a single \gls{gw}. Application-layer packets are generated
according to a Poisson Process with aggregate packet generation rate $\lambda$
[pck/s], and can be either confirmed or unconfirmed.

For tractability, we assume perfect orthogonality between different \glspl{sf},
i.e., only packets employing the same \gls{sf} can collide. In this case, one of
the two packets can survive if its received power is sufficiently higher than
that of the colliding packet (collisions with more than two packets happen with
negligible probability and are not considered). While the orthogonality
assumption has been shown to have an impact on the \gls{phy}-layer performance
of \gls{ul} only traffic~\cite{croce2018impact}, the results discussed in
Sec.~\ref{sec:results} show that the effect is much more limited in the presence
of confirmed traffic, where the performance is severely limited by other
factors.

\begin{figure}[t]
  \centering \begin{tikzpicture}

  \def\bottomy{0};
  \def\dashedsidelength{0.5};
  \def\blockheight{0.5};
  \def\blockwidth{7};
  \def\rcvtwoy{5.1};
  \def\rcvoney{3.3};
  \def\inty{1.4};

  \def\filteredx{1.3};
  \def\unconfx{2.5};
  \def\firstx{3.7};
  \def\secondx{5};
  \def\thirdx{6};
  \def\basey{0.5};
  \def\unconfy{2.5};
  \def\firsty{2.5};
  \def\secondy{4.3};
  \def\arrowl{0.6};

  \def\ratelabelsx{7};

  \def\firstonduration{0.8};
  \def\firstoffduration{3};
  \def\secondonduration{1.8};
  \def\secondoffduration{1.8};

  \draw (\dashedsidelength,\rcvtwoy) rectangle node[above, yshift=0.2cm]{\footnotesize{SB2 process}} ({\blockwidth + \dashedsidelength},{\rcvtwoy + \blockheight});
  \draw [dashed] (0,\rcvtwoy) -- (\dashedsidelength,\rcvtwoy);
  \draw [dashed] (0,{\rcvtwoy + \blockheight}) -- (\dashedsidelength,{\rcvtwoy + \blockheight});
  \draw [dashed] ({\blockwidth + \dashedsidelength},\rcvtwoy) -- ({\blockwidth + 2*\dashedsidelength},\rcvtwoy);
  \draw [dashed] ({\blockwidth + \dashedsidelength},{\rcvtwoy + \blockheight}) -- ({\blockwidth + 2*\dashedsidelength},{\rcvtwoy + \blockheight});

  \draw [pattern=north west lines] (\dashedsidelength, \rcvtwoy) rectangle node{\contour{white}{\footnotesize{OFF}}} ({\secondx - \secondonduration}, {\rcvtwoy + \blockheight});
  \draw ({\secondx - \secondonduration}, \rcvtwoy) rectangle node{\footnotesize{ON}} (\secondx, {\rcvtwoy + \blockheight});
  \draw [pattern=north west lines] (\secondx, \rcvtwoy) rectangle node{\contour{white}{\footnotesize{OFF}}} ({\secondx + \secondoffduration}, {\rcvtwoy + \blockheight});
  \draw ({\secondx + \secondoffduration}, \rcvtwoy) rectangle node{\footnotesize{ON}} ({\blockwidth + \dashedsidelength}, {\rcvtwoy + \blockheight});

  \draw [white] ({\dashedsidelength},\rcvtwoy) -- ({\dashedsidelength},{\rcvtwoy + \blockheight});
  \draw [white] ({\dashedsidelength + \blockwidth},\rcvtwoy) -- ({\dashedsidelength + \blockwidth},{\rcvtwoy + \blockheight});

  \draw (\dashedsidelength,\rcvoney) rectangle node[above, yshift=0.2cm]{\footnotesize{SB1 process}} ({\blockwidth + \dashedsidelength},{\rcvoney + \blockheight});
  \draw [dashed] (0,\rcvoney) -- (\dashedsidelength,\rcvoney);
  \draw [dashed] (0,{\rcvoney + \blockheight}) -- (\dashedsidelength,{\rcvoney + \blockheight});
  \draw [dashed] ({\blockwidth + \dashedsidelength},\rcvoney) -- ({\blockwidth + 2*\dashedsidelength},\rcvoney);
  \draw [dashed] ({\blockwidth + \dashedsidelength},{\rcvoney + \blockheight}) -- ({\blockwidth + 2*\dashedsidelength},{\rcvoney + \blockheight});

  \draw ({\dashedsidelength}, \rcvoney) rectangle node{\footnotesize{ON}} (\firstx, {\rcvoney + \blockheight});
  \draw [pattern=north west lines] (\firstx, \rcvoney) rectangle node{\contour{white}{\footnotesize{OFF}}} ({\firstx + \firstoffduration}, {\rcvoney + \blockheight});
  \draw ({\firstx + \firstoffduration}, \rcvoney) rectangle node{\footnotesize{ON}} ({\blockwidth + \dashedsidelength}, {\rcvoney + \blockheight});

  \draw [white] ({\dashedsidelength},\rcvoney) -- ({\dashedsidelength},{\rcvoney + \blockheight});
  \draw [white] ({\dashedsidelength + \blockwidth},\rcvoney) -- ({\dashedsidelength + \blockwidth},{\rcvoney + \blockheight});

  \draw [pattern=crosshatch] (\dashedsidelength,\inty) rectangle node[above, yshift=0.2cm]{\footnotesize{Interference, no demodulators and GW in TX filter}} ({\blockwidth + \dashedsidelength},{\inty + \blockheight});
  \draw [dashed] (0,\inty) -- (\dashedsidelength,\inty);
  \draw [dashed] (0,{\inty + \blockheight}) -- (\dashedsidelength,{\inty + \blockheight});
  \draw [dashed] ({\blockwidth + \dashedsidelength},\inty) -- ({\blockwidth + 2*\dashedsidelength},\inty);
  \draw [dashed] ({\blockwidth + \dashedsidelength},{\inty + \blockheight}) -- ({\blockwidth + 2*\dashedsidelength},{\inty + \blockheight});

  \draw [white] ({\dashedsidelength},\inty) -- ({\dashedsidelength},{\inty + \blockheight});
  \draw [white] ({\dashedsidelength + \blockwidth},\inty) -- ({\dashedsidelength + \blockwidth},{\inty + \blockheight});

  \draw[thick,->] (\unconfx,\basey) -- (\unconfx,{\basey + \arrowl}) node[midway, xshift=-0.7em] {\footnotesize{B}};
  \draw[thick,->] (\unconfx,\unconfy) -- (\unconfx,{\unconfy + \arrowl});

  \draw[thick,->] (\firstx,\basey) -- (\firstx,{\basey + \arrowl}) node[midway, xshift=-0.7em] {\footnotesize{C}};
  \draw[thick,->] (\firstx,\firsty) -- (\firstx,{\firsty + \arrowl});

  \draw[thick,->] (\secondx,\basey) -- (\secondx,{\basey + \arrowl}) node[midway, xshift=-0.7em] {\footnotesize{D}};
  \draw[thick,->] (\secondx,\firsty) -- (\secondx,{\firsty + \arrowl});
  \draw[thick,->] (\secondx,\secondy) -- (\secondx,{\secondy + \arrowl});

  \draw[thick,->] (\thirdx,\basey) -- (\thirdx,{\basey + \arrowl}) node[midway, xshift=-0.7em] {\footnotesize{E}};
  \draw[thick,->] (\thirdx,\firsty) -- (\thirdx,{\firsty + \arrowl});
  \draw[thick,->] (\thirdx,\secondy) -- (\thirdx,{\secondy + \arrowl});

  \draw[thick,->] (\filteredx,\basey) -- (\filteredx,{\basey + \arrowl}) node[midway, xshift=-0.7em] {\footnotesize{A}};

  \draw[thick,->] (0,0) -- ({2 * \dashedsidelength + \blockwidth},0) node[anchor=north west]{time};

  \node[] at (\ratelabelsx,{\basey + \arrowl / 2}) {$R^{phy}$};
  \node[] at (\ratelabelsx,{\firsty + \arrowl / 2}) {$r^1$};
  \node[] at (\ratelabelsx,{\secondy + \arrowl / 2}) {$r^2$};

\end{tikzpicture}
  \caption{Representation of the model's packet filtering structure. $R^{phy}$ is
    the rate of \gls{ul} traffic (see~\eqref{eq:rphytot}), while $r^1$ and
    $r^2$ represent the rate of \glspl{ack} sent in \gls{sb1} and \gls{sb2},
    respectively (see~\eqref{eq:r1},~\eqref{eq:r2}).}
  \label{fig:diagram}
  \vspace{-1em}
\end{figure}
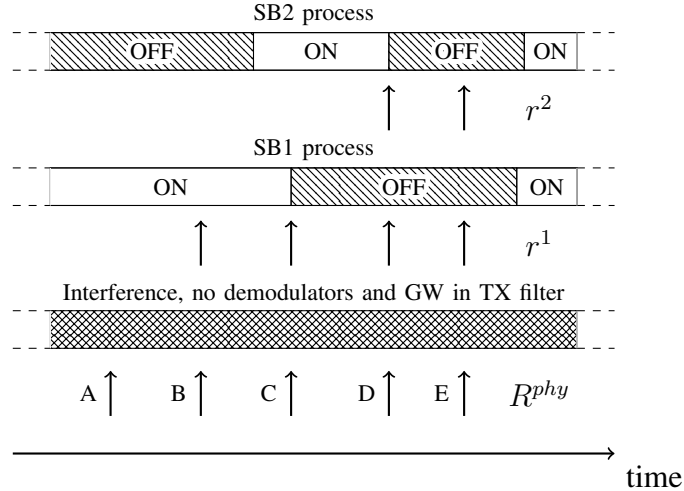

Fig.~\ref{fig:diagram} shows the structure of the packet reception model,
consisting in successive filtering of Poisson processes. At the base of the
figure, arrows are used to represent the \gls{ul} traffic generated by the
\glspl{ed}, including both new packet transmissions and re-transmissions of
failed packets. This process is assumed to be Poisson for tractability,
ignoring the fact that re-transmissions of a certain packet are correlated
in time because of \gls{dc} limitations. An initial filtering of this
process excludes some arrivals, modeling packet losses due to interference
from other \glspl{ed}, unavailability of \gls{gw} demodulators, or ongoing
\gls{dl} transmissions from the \gls{gw}. This yields a process with a
reduced rate, which now represents the packets that are correctly received
by the \gls{gw}.

When the received \gls{ul} message requires confirmation, an \gls{ack} must be
sent by the \gls{gw} during one of the two receive windows of the target
\gls{ed}. The ability of the \gls{gw} to perform such a transmission is modeled
through two independent alternating renewal processes, in which the system
alternates between the ON and OFF states. The two processes represent the
opportunity of sending the \gls{ack} in \gls{rx1} or \gls{rx2}, respectively,
which are opened on \gls{sb1} or \gls{sb2}, i.e., on the shared or dedicated
sub-band. If a confirmed packet finds a process in the ON state, it means that
the \gls{gw} will be able to send an \gls{ack} in that sub-band. In this case,
the process will switch to the OFF state to model the unavailability of that
sub-channel for a certain period of time following the ACK transmission, due to
the DC restrictions.

Since the sub-bands are disjoint, we assume that the two processes are
uncorrelated, neglecting the fact that the very packets that need to be served
in \gls{sb2} are those that found \gls{sb1} in the OFF state. If
the \gls{dl} packet finds at least one of the two processes in the ON state, an
\gls{ack} is sent. If the \gls{ack} is sent on \gls{sb1} (hence, using frequencies
shared by \gls{ul} and \gls{dl} traffic), it can be destroyed by the interference created by other \glspl{ed}.
If the \gls{ack} is sent on \gls{sb2}, instead, it is assumed to be
always successful.

For the sake of clarity, the following list describes some examples of the
life cycle of the packets in Fig.~\ref{fig:diagram}:
\begin{enumerate}[(A)]
\item This packet is lost because of interference or \gls{gw} transmission
  or unavailability of demodulators. Hence, it does not pass the first
  filter.
\item This is an unconfirmed \gls{ul} packet, which is successfully
  received by the \gls{gw}. It does not generate any \gls{ack}.
\item This is a confirmed packet successfully received by the \gls{gw}. It
  generates an \gls{ack}, which finds the \gls{sb1} process in the ON state. The
  \gls{ack} is successfully sent, and the \gls{sb1} process switches to the
  OFF state.
\item This is another confirmed packet which is successfully received by
  the \gls{gw}. Since the \gls{gw} has just sent an \gls{ack} for packet
  (C), it cannot reply in \gls{sb1} due to \gls{dc} constraints; \gls{sb2}
  is however in the ON state, and the \gls{gw} can thus reply to the
  \gls{ed}, making the second process switch to the OFF state.
\item This is another confirmed packet, which gets a treatment similar to
  that of packet (D). However, since the \gls{gw} has transmitted the
  \gls{ack} for packet (D) and is still under the \gls{dc} constraints, it
  cannot reply to packet (E) in either of the two receive windows (both
  \gls{sb1} and \gls{sb2} processes are in the OFF state). The \gls{dl}
  packet is hence discarded, and the \gls{ed} will re-transmit the \gls{ul}
  message at a later time.
\end{enumerate}

\subsection{Model Quantities}\label{sec:quantities}

Our model offers some tunable parameters to increase its flexibility,
enabling the evaluation of the network performance in various
configurations with minimal effort. The model makes it possible to specify
the following values:
\begin{itemize}
  \item $\mathcal{SF}=\{7,\ldots,12\}$ indicates the set of all SFs.
  \item $\alpha$: fraction of application-layer traffic requiring confirmation;
  \item $p^u_{i}, p^c_{i}$: fraction of devices generating unconfirmed and
    confirmed traffic with a specific SF $i \in \mathcal{SF}$, respectively.
    Note that $\sum_{i \in \mathcal{SF}} p_i^u = \sum_{i \in \mathcal{SF}} p_i^c
    = 1$;
  \item $h$: number of times an application-layer unconfirmed packet is
        transmitted;
      \item $m$: maximum number of transmission attempts for confirmed packets;
      \item $\delta_{SB1}$ and $\delta_{SB2}$: ratio between silent time and
        transmission time in SB$k$, corresponding to the \gls{dc} constraint.
        For instance, in Europe, we have $\delta_{SB1}=99$ and $\delta_{SB2}=9$
        corresponding to a \gls{dc} of 1\% in SB1 and 10\% in SB2. In general,
        when $\delta_{SBk}>0$ the \gls{dc} constraint applies to all devices
        transmitting in subchannel SB$k$. Instead, the setting $\delta_{SB_k}=0$
        corresponds to a DC constraint of 100\%, which means that there is no
        limitation on the transmission time\footnote{This setting is not allowed
          by current RF recommendations but is considered in this study to gain
          insights on the impact of \gls{dc} limitations in the considered
          scenarios.};
      \item $\tau_1$ and $\tau_2$: prioritization flags. If $\tau_k=0$, the
        \gls{gw} prioritizes reception operations over transmission during the
        $k$-th receive window, with $k=1, 2$. In this case, the \gls{gw} will
        drop any DL message that needs to be transmitted while a \gls{ul}
        reception is ongoing. Instead, if TX is prioritized ($\tau_k = 1$), the
        reception of any incoming packet will be interrupted in order to send
        the \gls{ack};
  \item $C$: number of \gls{ul} frequency channels. Note that each \gls{ul}
        channel can also be used for \gls{dl} transmissions. Instead, the
        channel in \gls{sb2} is \gls{dl} only;
  \item $T_{i}^{ack_2}$: duration of the transmission of the \gls{ack} in
        \gls{rx2} when using \gls{sf} $i$. (The standard requires the use of SF
        12 in \gls{rx2} as a pre-configured setting, corresponding to
        $T_{12}^{ack_2}$. Note that this default setting can be changed by the
        NS, and accordingly in our model.)
      \item $T_i^{data}$ and $T_i^{ack_1}$ indicate the time durations of a data
        packet and of an \gls{ack} transmitted in \gls{sb1} with \gls{sf} $i$,
        respectively. If \glspl{ack} transmitted in \gls{sb2} use \gls{sf}12,
        irrespective of the \gls{sf} employed in the \gls{ul} transmission, then
        $T_i^{ack_2} = T_{12}^{ack_1}, \: \forall i \in \mathcal{SF}$.
\end{itemize}

In the formulas, the notation generally respects the following scheme. The
probability is indicated with $S$ or $F$ if it corresponds to a ``success''
or ``failure'' event, respectively; if this rule does not apply, the probability
is denoted simply as $P$. The superscript indicates the considered event, while
the subscript the \gls{sf}. For example, in~\eqref{eq:Sint}, the symbol
$S_i^{INT}$ represents the probability of successfully surviving interference
when using \gls{sf}~$i$. Different uses of the notation are specified in the
text. The following sections provide a mathematical formulation for some
relevant quantities in this model.

\subsection{Uplink traffic rates}
\label{sec:ultrafficrates}

The assumption of perfect orthogonality between different \glspl{sf} makes it
possible to split the network traffic in different logical channels that do not
interfere with each other. The traffic load for each \gls{sf} $i$ is split
uniformly over the given $C$ frequency channels (since \glspl{ed} pick a random
\gls{ul} frequency for each transmission attempt). Thus, the traffic generated
at the application layer by the \glspl{ed} using confirmed and unconfirmed
messages is, respectively, given by:
\begin{align}
  \label{eq:RateApp}
  R_i^{c, app} &= \frac{p_i^c \cdot \lambda}{C} \cdot \alpha, \\
  R_i^{u, app} &= \frac{p_i^u \cdot \lambda}{C} \cdot (1 - \alpha).
\end{align}

Since \glspl{ed} using unconfirmed traffic will perform $h$ transmissions of
each application-layer packet, the PHY rate of these devices can be computed as
$R_i^{u, phy} = R_i^{u, app} \cdot h$. For \glspl{ed} transmitting confirmed
messages, instead, the number of re-transmitted packets depends on the success
of both the \gls{ul} transmission and the corresponding \gls{ack}. We indicate
as $P_{i,j}^{DL}$ the probability that a confirmed \gls{ul} packet sent with
\gls{sf} $i$ is successfully received and acknowledged at the $j$-th
transmission attempt, which will be derived in~\eqref{eq:psucc_dl}.
Therefore, we have that the rate of confirmed packets transmitted at \gls{sf}
$i$, $R_i^{c,phy}$, is given by the product of the application-level rate,
$R_i^{c,app}$, and the average number of times a confirmed packet is transmitted
at the \gls{phy} layer.

\begin{align}
  \label{eq:Rc}
  \begin{split}
    R_i^{c, phy} = R_i^{c, app} \Bigg[ &\sum_{j = 1}^{m-1} j \cdot
    P_{i,j}^{DL} 
    + m \left( 1 - \sum_{j = 1}^{m-1}P_{i,j}^{DL} \right) \Bigg].
  \end{split}
\end{align}
The first summation in the square brackets of~\eqref{eq:Rc} takes into account
transmissions that are successfully received before the $m$th attempt, while the
second term considers the case when the packet is transmitted $m$ times
(irrespective of whether the last transmission is successful or not).

The total traffic for a single frequency channel and for \gls{sf}~$i$ is
therefore given by
\begin{equation}
  \label{eq:rphytot}
  R_i^{phy} = R_i^{u, phy} + R_i^{c, phy}.
\end{equation}

In general, the distribution of the \glspl{sf} for the transmitted packets at the
\gls{phy} layer will differ from the native distribution of \glspl{sf} among the
devices, \{$p_i^u, p_i^c$\}, because of re-transmissions. Thus, we
define
\begin{equation}
  \label{eq:d}
  d_i = \frac{R_i^{phy}}{\sum_j R_j^{phy}},
\end{equation}
as the ratio of \gls{phy} layer packets that are transmitted at \gls{sf} $i \in
\mathcal{SF}$.

\subsection{PHY layer probabilities}
\label{sec:phy-probs}

A \gls{ul} packet is successfully received by the \gls{gw} if all the following
conditions are met: (i) it does not overlap with another \gls{ul} transmission
using the same \gls{sf} on the same frequency, or it overlaps with another
\gls{ul} packet, but the received power is sufficiently large to allow for
correct decoding despite the interference (capture), (ii) it does not overlap
with a \gls{gw} \gls{dl} transmission in any channel, and (iii) it finds an
available demodulator. These conditions are represented by the first filter in
Fig.~\ref{fig:diagram}.

Since packets are generated following a Poisson process, the probability of
event (i) is given by two components. The first is the probability that there
are no other arrivals during the $2T_i^{data}$ vulnerability period across the
packet arrival instant. The second, considers a collision with one packet, and
the fact that the receiver successfully captures the frame. For the \gls{ul}, we
consider the capture probability $ \mathbb{W}^{GW}$ as computed
in~\cite{bankov2017mathem}. Since these two events are disjoint, the probability
of surviving interference (event (i)) is given by the sum of the two components,
which results in
\begin{equation}
  \label{eq:Sint}
  S_i^{INT} = e^{-2T_i^{data}R_i^{phy}} +
  2T_i^{data}R_i^{phy}e^{-2T_i^{data}R_i^{phy}} \cdot \mathbb{W}^{GW},
\end{equation}
where, in the right-most term, we computed the probability that either of the
two colliding packets is captured (collision events with more than two packets
are neglected).

To compute the probability of event (ii), we observe that a \gls{ul} message is
always lost when it arrives at the \gls{gw} during the transmission of an
\gls{ack}. Otherwise, the \gls{gw} will start the reception of the \gls{ul}
message, which will take a time $T_i^{data}$. If reception on SB$k$ is
prioritized (i.e., $\tau_k=0$), this process cannot be interrupted, and the
\gls{ul} message will be successfully delivered to the \gls{ns}. Conversely, if
$\tau_k=1$, i.e., we prioritize transmission on SB$k$, the reception of the
\gls{ul} packet may be aborted at any time during the period $T_i^{data}$, in
order to give priority on the \gls{ack} transmission. Therefore, the
vulnerability period is given by the \gls{ack} transmission time
$T_{s}^{ack_k}$, to which we need to add the interval $T_i^{data}$ only if
$\tau_k=1$. Denoting by $b_s^k$ the probability that an \gls{ack} is transmitted
on SB$k$ with \gls{sf} $s\in \mathcal{SF}$ (which will be derived later
in~\eqref{eq:bk}), the average vulnerability period is then given by
$\overline{T_k} = \sum_{s\in \mathcal{SF}} b_s^k T_s^{ack_k} + T_i^{data}\cdot
\tau_k$. Now, according to the Poisson Arrivals See Time Averages (PASTA)
property, the probability that a \gls{ul} packet arrival falls in the
vulnerability period of channel SB$k$, with $k=1,2$, can be expressed as
\begin{equation}
  \label{eq:f}
  F_{i}^{TXk}=\frac{\sum_{s\in \mathcal{SF}}{b_s^k T_s^{ack_k}}+ T_i^{data} \cdot \tau_k}{E_{ON}^k+E_{OFF}^k},
\end{equation}
where the denominator is the mean renewal time of the SB$k$ process, given
by the sum of $E_{ON}^k$ and $E_{OFF}^k$, i.e., the expected times the SB$k$
process spends in the ON and OFF states during a renewal period (ON-OFF cycle),
which will be computed in~\eqref{eq:eon} and~\eqref{eq:eoff}. Then, assuming
(for ease of analysis) that events in \gls{sb1} and \gls{sb2} are independent,
the probability that a \gls{ul} packet is successfully received (event (ii)) is
given by
\begin{equation}
  \label{eq:Sitx}
  S_i^{TX} =(1-F_i^{TX1})(1-F_i^{TX2}).
\end{equation}

Next, we compute the probability of event (iii), i.e., that at least one
demodulator out of 8 is available. Each demodulator chain is modeled through an
alternating renewal process, where the demodulator can be in an ``available''
state $A$, when idle or in a ``locked'' state $L$, when occupied with the
reception of another signal. We assume that the different demodulators are
activated in succession: if all are available, an incoming signal will be
received by the first demodulator; if the first demodulator is in the $L$ state,
the packet will be handled by the second demodulator, and so on. Let $E^L$ be
the expected time a demodulator will be locked on a incoming signal. Since the
occupation will last for the duration of \gls{ul} LoRa packets at the \gls{phy}
layer, we have:
\begin{equation}
  \label{eq:el}
  E^L = \sum_{i \in \mathcal{SF}}d_i \cdot T_i^{data}.
\end{equation}
The average time the first demodulator is in the $A$ state, instead, is
computed as the average inter-arrival time of \gls{ul} packets, regardless
of their \gls{sf} and selected frequency:
\begin{equation}
  E^{A, 1} = \frac{1}{C \cdot \sum_{i\in \mathcal{SF}} R_i^{phy}}.
\end{equation}
Then, the process of packets that require the second demodulator is filtered by
the probability of finding the first demodulator occupied. Thus, the expected
time the second demodulator is available is given by
\begin{equation}
  E^{A, 2} = \frac{E^{A,1}}{P^{L,1}} = \frac{1}{P^{L, 1} \cdot C \cdot \sum_{i\in \mathcal{SF}} R_i^{phy}},
\end{equation}
where $P^{L,1}$ is the probability that the first modulator is in the $L$ state
(see~\eqref{eq:pl1}).
With a similar reasoning, we compute the expected time for which the $j$-th
demodulator is available as
\begin{equation}
  E^{A, j} = \frac{E^{A,j-1}}{P^{L,j-1}} = \frac{1}{\prod_{\ell=1}^{j-1}P^{L, \ell} \cdot C \cdot \sum_{i\in \mathcal{SF}} R_i^{phy}}.
\end{equation}
The probability $P^{L,\ell}$ of finding the $\ell$-th demodulator in the $L$ state, in turn, can be expressed as
\begin{equation}
  \label{eq:pl1}
  P^{L, \ell} = \frac{E^{L}}{E^{A, \ell} + E^{L}}.
\end{equation}

Then, a packet finds an available demodulator (event (iii)) with probability:
\begin{equation}
  \label{eq:Sdemod}
  S^{demod} = 1 - \prod_{j=1}^8 P^{L, j}.
\end{equation}

The overall \gls{ul} packet success probability, considering events (i), (ii)
and (iii) described above, is finally expressed as
\begin{equation}
  \label{eq:Sul} S_i^{UL} = S_i^{INT} \cdot S_i^{TX} \cdot S^{demod} .
\end{equation}

\subsection{\gls{ack} transmission}
\label{sec:acktransmission}
Once a confirmed packet is correctly received by the \gls{gw}, an \gls{ack}
needs to be transmitted back to the \gls{ed}. Eq.~\eqref{eq:Sul} gives the
probability of successful packet reception at the \gls{gw}. Therefore, the rate
of \gls{ack} messages that the \gls{gw} will try to send in \gls{sb1} is:
\begin{equation}
  \label{eq:r1}
  r_i^1 = R_i^{c, phy} \cdot S_i^{UL}.
\end{equation}
\begin{figure}[t]
  \centering
  \usetikzlibrary{calc}
\begin{tikzpicture}[>=latex]

  %
  %
  \tikzstyle{state} = [draw, fill=white, rectangle, node distance=3em, minimum width=6em, minimum height=1.5em, font={\footnotesize}]
  \tikzstyle{failure} = [draw, fill=white, rectangle, node distance=3em, minimum width=4em, minimum height=1.5em, font={\footnotesize}]
  \tikzstyle{stateEdgePortion} = [black,thick];
  \tikzstyle{stateEdge} = [stateEdgePortion,->];
  \tikzstyle{edgeLabel} = [text centered, font={\sffamily\small}];

  %
  %
  \node[state, name=packetrx] {Successul reception of confirmed UL packet};
  \node[state, name=rx1avail, below of=packetrx] {SB1 is ON};
  \node[state, name=sharedch, below of=rx1avail] {Shared channel};

  \node[state, name=tryrx2, right of=packetrx, xshift=11em] {Try SB2};
  \node[state, name=rx2avail, below of=tryrx2] {SB2 is ON};
  \node[state, name=dedicatedch, below of=rx2avail] {Dedicated channel};
  \node[state, name=success, below of=dedicatedch, minimum width=20em, xshift=-7em, yshift=-1em] {Success};
  \node[failure, name=failure1, right of=tryrx2, xshift=8em] {Failure};
  \node[failure, name=failure2, right of=rx2avail, xshift=8em] {Failure};
  \node[failure, name=failure3, left of=sharedch, xshift=-8em] {Failure};


  %
  %
  \draw ($(packetrx.south)$)
    edge[stateEdge] node[edgeLabel, xshift=1.5em]{$P^{\rm ON1}$}
    ($(rx1avail.north)$);
  \draw ($(rx1avail.south)$)
    edge[stateEdge] node[edgeLabel, xshift=1.5em]{$P^{T1}$}
    ($(sharedch.north)$);
  \draw ($(sharedch.south)$)
    edge[stateEdge] node[edgeLabel, xshift=-3.5em]{$S_i^{\rm INT, ack1}$}
    ($(success.north)$);

  \draw ($(packetrx.east)$)
    edge[stateEdge] node[edgeLabel, yshift=1em]{$P^{\rm OFF1}$}
    ($(tryrx2.west)$);
  \draw ($(tryrx2.south)$)
    edge[stateEdge] node[edgeLabel, xshift=1.5em]{$P^{\rm ON2}$}
    ($(rx2avail.north)$);
  \draw ($(rx1avail.east)$)
    edge[stateEdge] node[edgeLabel, yshift=-1em, xshift=1em]{$1 - P^{T1}$}
    ($(tryrx2.west)$);
  \draw ($(rx2avail.south)$)
    edge[stateEdge] node[edgeLabel, xshift=1.5em]{$P^{T2}$}
    ($(dedicatedch.north)$);
  \draw ($(dedicatedch.south)$)
    edge[stateEdge] node[edgeLabel, xshift=1.5em]{$1$}
    ($(success.north)$);

  \draw [dashed] ($(tryrx2.east)$) edge[stateEdge] node[edgeLabel, yshift=1em]{$1 - P^{\rm ON2}$} ($(failure1.west)$);
  \draw [dashed] ($(rx2avail.east)$) edge[stateEdge] node[edgeLabel, yshift=1em]{$1 - P^{T2}$} ($(failure2.west)$);
  \draw [dashed] ($(sharedch.west)$) edge[stateEdge] node[edgeLabel, yshift=1em]{$1-S_i^{\rm INT, ack1}$} ($(failure3.east)$);

\end{tikzpicture}
  \caption{Diagram for successful \gls{ack} reception.}
  \label{fig:dldiagram}
  \vspace{-1em}
\end{figure}
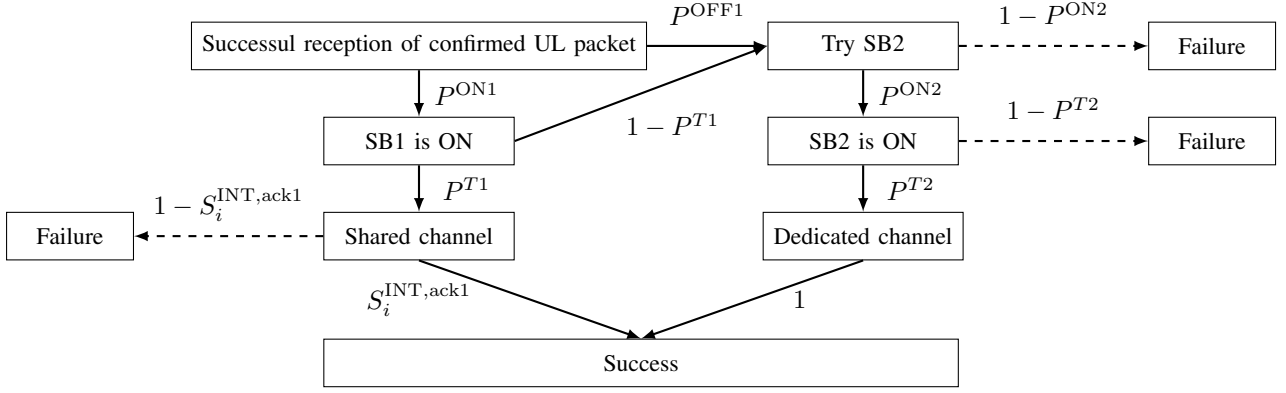
A visual representation of the possible \gls{ack} life cycles considered in the
model is shown in Fig.~\ref{fig:dldiagram}. Labels refer to the probabilities of
the different events, which we derive next. In general, an \gls{ack} is
transmitted in SB$k$ if both the following conditions hold: (i) $\tau_k=1$ (TX
is prioritized) or $\tau_k=0$ and the GW is idle; (ii) SB$k$ is available (i.e.,
not blocked by DC constraints). If either condition is not satisfied, the
\gls{ack} is dropped.

Let $T$ denote the event ``the \gls{gw} \textit{may} transmit,''
which depends on the TX/RX prioritization policy. If $\tau_k=1$, the
\gls{gw} can transmit the \gls{dl} packet whenever it needs to; otherwise,
if $\tau_k=0$, the \gls{gw} can transmit in SB$k$ only if no reception is
ongoing. We denote by $P^{T,k}$ the probability of $T$, which can be computed as
\begin{equation}
  \label{eq:Pnorx}
  P^{T, k} =
  \begin{cases}
    1, & \textrm{if $\tau_k$ = 1}; \\
    e^{-\sum_{i \in \mathcal{SF}} C \cdot R_i^{phy} T_i^{data}}, &
    \textrm{if $\tau_k$ = 0};
  \end{cases}
\end{equation}
where the second expression is the probability that no \gls{ul} packet was
generated in the last $T^{data}_i$ seconds.

If \gls{sb1} is not available, the \gls{gw} will try to process the
\gls{ack} in \gls{sb2}. Such packets form a process with rate
\begin{equation}
  \label{eq:r2}
  r_i^2= r_i^1 [P^{OFF, 1} + P^{ON, 1}(1 - P^{T, 1})],
\end{equation}
where $P^{ON, 1}$ and $P^{OFF, 1}$ are the probabilities of finding \gls{sb1}
in the ON and OFF state, respectively, and $(1 - P^{T, 1})$ is the
probability that the \gls{gw} is not available for \gls{dl} transmission.
The ON and OFF probabilities for the SB$k$ process, with $k=1,2$, are given
by
\begin{align}
  \label{eq:Ponoff}
  P^{ON, k}  &= \frac{E^{ON, k}}{E^{ON, k} + E^{OFF ,k}}, \\
  P^{OFF, k} &= \frac{E^{OFF, k}}{E^{ON, k} + E^{OFF, k}},
\end{align}
where $E^{ON,k}$ and $E^{OFF,k}$ are the mean sojourn times in ON and OFF
states, respectively, which are computed as follows.
By considering the arrival rate of successful \gls{ul} packets in the
$k$-th sub-band, we have:
\begin{align}
  \label{eq:eon}
  \begin{split}
    E^{\mathrm{ON}k} &= \frac{1}{\sum_{i \in \mathcal{SF}} C \cdot r_i^k}.
  \end{split}
\end{align}
Note that the switch from the ON to the OFF state will be caused by a packet
sent in any of the $C$ \gls{ul} channels: therefore, we need to multiply the
rates $r_i^k$ of arrivals to SB$k$ with SF $i$ by the number of available
channels.

In order to compute the expected duration of the OFF periods, we first need
to derive the probability distribution $b_i^k$ of the \glspl{sf} used for
\gls{ack} transmissions, which is given by
\begin{equation}
  \label{eq:bk}
  b_i^k = \frac{r_i^k}{\sum\limits_{s \in \mathcal{SF}}r_s^k}.
\end{equation}
%
%
In our model, the OFF period accounts for the time the \gls{gw} is
prevented from transmitting a new data packet, which includes the time to
send the ACK using the given \gls{sf}, plus the waiting time imposed by the
\gls{dc} limitations. We hence have
\begin{align}
  \label{eq:eoff}
  \begin{split}
    E^{\rm OFF, 1} &= \sum_{s \in \mathcal{SF}} b_s^1 (T_s^{\rm ack_1} + \delta_{SB1} \cdot T_s^{\rm ack_1}), \\
    E^{\rm OFF, 2} &= \sum_{s \in \mathcal{SF}} b_s^2 (T_s^{\rm ack_2} + \delta_{SB2} \cdot
    T_s^{\rm ack_2}).
  \end{split}
\end{align}
(Note that, by including the parameter $\delta_{SBk}$ as defined in
Sec.~\ref{sec:scenario}, we can change the \gls{dc} limitations in the $k$-th
sub-band, thus making it possible to analyze its impact.)

Finally, we remark that \gls{dl} packets sent by the \gls{gw} in \gls{sb1} also
have to avoid interference from other \glspl{ed}. In the absence of collisions,
the vulnerability period is given by the sum of two terms. The first term
corresponds to the case of no \gls{ul} transmissions starting while the \gls{dl}
packet is being sent ($T^{ack_1}$); the second term represents the event where
no \gls{ul} transmissions started before the \gls{ack} is sent. Note that if
$\tau_1=0$ the second term is not present, since in that case the \gls{ack}
would not be generated at all. Furthermore, an \gls{ack} can survive an
interfering packet sent by another \gls{ed} in case of capture, which happens
with probability $\mathbb{W}^{ED}$ (equivalent to the $\mathbb{W}^{Mote}$ as
derived in~\cite{bankov2017mathem}). Therefore, the probability that the
\gls{ack} does not collide with a \gls{ul} packet in \gls{sb1}, or is captured
despite the collision, is equal to
\begin{equation}
  \label{eq:SintAck}
  S_i^{INT, ack_1} =
  e^{-R_i^{phy} (T_i^{ack_1} + \tau_1 \cdot T_i^{data})} +
  R_i^{phy} (T_i^{ack_1} + T_i^{data}) \cdot e^{-R_i^{phy} (T_i^{ack_1} + T_i^{data})} \cdot \mathbb{W}^{ED}.
\end{equation}
For packets sent in \gls{sb2}, instead, the reception is assumed to
be always successful, since the 869.525~MHz channel is dedicated to \gls{dl}
communication and the \gls{gw} only transmits one packet at a time (note
that this assumption does not hold in the case of multiple \glspl{gw}).

\subsection{\gls{dl} success probability}
\label{sec:succprobs}
Given that a confirmed \gls{ul} packet sent with \gls{sf} $i$ has been
successfully received by the \gls{gw}, the probability that the corresponding
\gls{ack} is also successfully returned to the \gls{ed} is expressed as
\begin{equation}
  \label{eq:Sdl1ack} S_i^{\rm DL} = S_{i}^{\textrm{SB1}} +  S^{\textrm{SB2}},
\end{equation}
where $S_{i}^{\textrm{SB1}} $ describes the probability of a successful
\gls{ack} transmission in \gls{sb1} with \gls{sf} $i$, while $S^{\textrm{SB2}}$
accounts for the probability that \gls{sb1} is not available, and the \gls{ack}
is successfully sent in \gls{sb2}. These probabilities, in turn, can be
expressed as follows:
\begin{align}
  \label{eq:SB1B2}
  S_{i}^{\textrm{SB1}} &=\; P^{ON, 1} \cdot P^{T, 1} \cdot S_i^{INT, ack_1}, \\
  S^{\textrm{SB2}} &=\;[P^{OFF, 1} + P^{ON, 1} \cdot (1 - P^{T, 1}) ]\cdot P^{ON, 2} \cdot P^{T, 2}.
\end{align}

Fig.~\ref{fig:dldiagram} can be used as a reference for the computation of
this quantity.


Finally, we can compute the success probabilities over $m$ transmissions. We
recall that, for the sake of simplicity, we neglect the time correlation of
packet re-transmissions due to \gls{dc} constraints, (the impact of this
approximation will be analyzed by simulation). We recall that $P_{i,j}^{UL}$
indicates the probability that a \gls{ul} packet with \gls{sf} $i$ is
successfully received at the \gls{gw} at exactly the $j$-th transmission
attempt, which can be computed as:
\begin{equation}
  \label{eq:psucc_ul}
  P_{i,j}^{UL} = S_i^{UL} \left(1 - S_i^{UL}\right)^{j - 1} .
\end{equation}
Then, the \gls{ed} successfully receives the \gls{ack} at exactly the $j$-th
attempt if both the \gls{ul} and the \gls{dl} transmissions succeed. The
probability $P_{i,j}^{DL}$ of this event is hence given by:
\begin{equation}
  \label{eq:psucc_dl}
  P_{i,j}^{DL} = \left[1 - (S_i^{UL}S_i^{DL})\right]^{j - 1} \cdot (S_i^{UL}S_i^{DL}).
\end{equation}

Once all intermediate quantities are computed, the model can be summarized
by two inter-dependent equations:
\begin{equation*}
  \begin{cases*}
    S^{UL} = f(S^{UL}, S^{DL}),\\
    S^{DL} = g(S^{UL}, S^{DL}).
  \end{cases*}
\end{equation*}
where $S^{UL}=[S^{UL}_7,\ldots,S^{UL}_{12}]$ and
$S^{DL}=[S^{DL}_7,\ldots,S^{DL}_{12}]$, while $f()$ and $g()$ are implicit
functions given by the chaining of the sequence of operations that
yield~\eqref{eq:Sul} and~\eqref{eq:Sdl1ack}, respectively.

This system admits a fixed-point solution, which can be found through
fixed-point iteration. From a practical perspective, when initialized with the
states $S^{UL} = S^{DL} = [1, 1, 1, 1, 1, 1]$, the iterative process has always
reached convergence to the stable fixed point after a few iterations (order of
few units) for all the parameter combinations considered in this work. The proof
of the system's convergence is provided in~\cite{magrin2021proof}. An
implementation of the model, allowing the interested readers to easily replicate
the results shown in this paper, is publicly available
at~\cite{publishedmodelcode}.

\subsection{Performance metrics}
\label{sec:metrics}
To evaluate the system performance, we consider three classes of key performance
Sndicators, namely: reliability, delay, and fairness metrics which are better
detailed in the remainder of this section together with the methodology to
determine their value using the proposed model. Once a set of parameters is
fixed, the model can be solved and the performance metrics can be estimated
starting from $S^{UL}$ and $S^{DL}$. Conversely, it is possible to employ the
model to optimize a given performance metric, finding the parameter setting that
maximizes it, as shown in Sec.~\ref{sec:results}.

\subsubsection{Reliability Metrics}

We consider three \gls{pdr} indexes, namely:
\begin{itemize}
\item \textit{\gls{uu}}: fraction of (application-layer) unconfirmed
  packets that are successfully received by the \gls{gw};
\item \textit{\gls{cu}}: fraction of (application-layer) confirmed packets that
  are successfully received by the \gls{gw}, irrespective of whether or not the
  corresponding \gls{ack} is successfully returned to the \gls{ed};
\item \textit{\gls{cd}}: fraction of (application-layer) confirmed packets
  that are successfully acknowledged by the NS.
\end{itemize}
Clearly, CD $\leq$ CU, since a packet needs to be successfully received by
the \gls{gw} in order to be acknowledged. Note that the \gls{cu} metric
captures the performance of applications for which it is important to
deliver packets to the \gls{ns} and \glspl{ack} are only used to stop
re-transmissions (and thus avoid a useless increase in traffic), while
\gls{cd} is more interesting for applications that require the \glspl{ed}
to get explicit feedback from the \gls{ns}, for instance containing control
information addressed to the \gls{ed}.



We obtain the \gls{uu} and \gls{cu} values by averaging the \gls{ul} success
probability ($UU_i$ and $CU_i$ for unconfirmed and confirmed packets,
respectively) for each  \gls{sf} $i$ over the \gls{sf} distribution, i.e.,
\begin{equation}
  \label{eq:uu}
  {\rm UU} = \sum_{i \in \mathcal{SF}} \left(p_i^u \cdot \mathrm{UU}_i\right) = \sum_{i \in \mathcal{SF}} \left(p_i^u \cdot \sum\limits_{j=1}^h P_{i, j}^{UL}\right),
\end{equation}
\begin{equation}
  \label{eq:cu}
  {\rm CU} = \sum_{i \in \mathcal{SF}} \left(p_i^c \cdot \mathrm{CU}_i \right) = \sum_{i \in \mathcal{SF}} \left(p_i^c \cdot \sum\limits_{j=1}^mP_{i, j}^{UL}\right).
\end{equation}

Similarly, \gls{cd} is computed as the probability of success for a
confirmed packet within the available re-transmission attempts
\begin{equation}
  \label{eq:cd}
  {\rm CD} = \sum\limits_{i \in \mathcal{SF}} \left( p_i^c \cdot \sum\limits_{j=1}^mP^{DL}_{i,j}\right).
\end{equation}

\subsubsection{Delay Metrics}
We define two delay metrics, considering confirmed traffic only: $\Delta^{\rm
  UL}$ measures the time from the first transmission attempt to the successful
delivery to the \gls{gw} of an \gls{ul} confirmed packet, while $\Delta^{\rm
  DL}$ accounts for the time from the first transmission of a confirmed packet
to the successful reception of the corresponding reply. Delays are computed for
successful packets only, and the propagation delay is assumed to be negligible.
To compute these metrics with our model, we assume the RETRANSMIT\_TIMEOUT value
to be a uniformly distributed random variable with mean $\mu$, and consider that
\glspl{ed} employ the shared sub-band with $\delta_{SB1}$ \gls{dc} limitations.
Therefore, the average time between two transmissions of the same MAC-layer
packet by a device is given by:
\begin{equation}
  \label{eq:intertranmissionTime}
  \gamma_i = (\delta_{SB1} + 1) \cdot T_i^{data} + \mu.
\end{equation}

The average delay from the successful reception of a packet at the \gls{gw}
to the transmission of the \gls{ack} is given by:
\begin{equation}
  \label{eq:avgAckTransmissionTime}
  \phi_i = S_i^{\rm SB1} \cdot (1 + T_i^{ack_1}) + S^{\rm SB2} \cdot (2 + T_i^{ack_2}),
\end{equation}
where we take into account that the \gls{ack} will be served in SB1 (opened
after 1 second) with probability $S_i^{\rm SB1}$, and in SB2 (opened after 2
seconds) with probability $S^{\rm SB2}$.

If a packet is re-transmitted $m$ times, each re-transmission $j$ is
associated with a certain \gls{ul} success probability $P_{i,j}^{\rm UL}$.
The average delay at each \gls{sf} $i \in \mathcal{SF}$ can be computed as:
\begin{equation}
  \label{eq:uldelay}
  \Delta^{\rm UL} = \sum_{i\in\mathcal{SF}} p_i^c \cdot \left(  \sum_{j=1}^m \bar{P}_{i,j}^{\rm UL} \left(T_i^{data} + (j-1) \cdot \gamma_i\right)\right),
\end{equation}
where we define $\bar{P}_{i,j}^{\rm UL} = P_{i,j}^{\rm
  UL}/\sum_jP_{i,j}^{\rm UL}$ to obtain the distribution of successful
\gls{ul} packet transmissions.

Similarly, we can compute the average \gls{ack} delay:
\begin{equation}
  \label{eq:dldelay}
  \Delta^{\rm DL} = \sum_{i\in\mathcal{SF}} p_i^c \cdot \left( \sum_{j=1}^m \bar{P}_{i,j}^{\rm DL} \left(T_i^{data} + (j-1) \cdot \gamma_i + j \cdot \phi_i\right) \right),
\end{equation}
where, in addition to the inter-transmission time between two packets, we
also account for the time to perform the \gls{ack} transmission.

\subsubsection{Fairness}
Finally, we consider the fairness of the system in different scenarios. Indeed,
\glspl{ed} employing confirmed traffic or higher \glspl{sf} will use more
system resources (e.g., channel occupancy), possibly affecting the application
performance of devices that employ different settings.
To this aim, we use Jain's fairness index, defined as
\begin{equation}
  \label{eq:j}
  J(\mathbf{x}) = \frac{\Big(\sum_{i=1}^{n}x_i\Big)^2}{n \cdot \sum_{i=1}^{n}x_i^2},
\end{equation}
where $n$ is the total number of user categories, each with throughput
$x_i$. Note that $1/n \leq J(\mathbf{x}) \leq 1$, and the system is perfectly fair if
$J(\mathbf{x}) = 1$. In particular, in the following section, we will consider
the fairness among devices employing different \glspl{sf}. Furthermore, since
all the devices have equal packet generation rate, and transmit packets with the same length, instead of the throughput
we can simply consider the \gls{ul} success probability, i.e., \gls{uu} for
nodes employing unconfirmed traffic and \gls{cu} for devices transmitting confirmed
messages. Therefore, the fairness is computed by taking
$\mathbf{x} = [\mathbf{x^u}, \mathbf{x^c}]$, where the elements correspond to
$x_i^u = UU_i$, and $x_i^c = CU_i$, as defined in~\eqref{eq:uu},~\eqref{eq:cu}.

\section{Network Simulations}
\label{sec:simulation}

In order to validate our model, we compared the performance estimates obtained
from the model with those observed in more realistic simulations, in which most
of the simplifying assumptions of the model are removed.

This section describes how we employ the LoRaWAN ns-3 module described
in~\cite{magrin2020thorough} to perform such a validation. To be noted that the
more accurate modeling of the LoRaWAN standard considered in the simulator comes
at the cost of a much larger computational time to assess the system
performance. Indeed, for the same parameter set, the performance evaluation is
basically instantaneous when employing the theoretical model, while each ns-3
simulation run takes in the order of tens of seconds, with execution times
rapidly increasing when the traffic load, the number of devices and the number
of required randomized runs grow.

The merit of the simulator is that it strives to be as realistic as possible,
also taking into account some factors that are overlooked by the model for
tractability reasons. For instance, the assumption of perfect orthogonality
between transmissions employing different \glspl{sf} is removed, and the
simulator relies on the link-level model provided
in~\cite{goursaud2015dedicated} to determine the actual reception probability in
case of overlapping transmissions, which also accounts for the capture effect.

\begin{table}[t]
  \footnotesize
  \centering
  \caption{Values of $T^{data}$, $T^{ack}$ and SF distributions $p$.
     Payload of data packets is 10 bytes; \glspl{ack} have no
    payload.}
  \begin{tabular}[c]{ccccc}
    \toprule
    SF & $T^{data}$ [s] & $T^{ack}$ [s] & $p_{\rm equal}$ & $p_{\rm EXPLoRa}$\\
    \midrule
    7  & 0.051 & 0.041 & 0.166 & 0.487 \\
    8  & 0.102 & 0.072 & 0.166 & 0.243 \\
    9  & 0.185 & 0.144 & 0.166 & 0.135 \\
    10 & 0.329 & 0.247 & 0.166 & 0.076 \\
    11 & 0.659 & 0.495 & 0.166 & 0.038 \\
    12 & 1.318 & 0.991 & 0.166 & 0.019 \\
    \bottomrule
  \end{tabular}
  \label{tab:resparams}
\end{table}

The simulation setting is as follows.
\begin{itemize}
\item \textit{Traffic load} -- The number of \glspl{ed} is fixed to 1200, and the \glspl{ed}' application layer is set to periodically generate packets to be transmitted by the MAC layer. The traffic load in the network is modified by varying the packet generation
  period. It is to be noted that this periodic traffic generation pattern is likely more realistic than the Poisson traffic assumed in the model. Nonetheless, the good match of simulation and analytical results confirms that the Poisson assumption is valid when the number of nodes is sufficiently large.

\item \textit{Channel allocation} -- We consider the typical frequency
  allocation scheme for Europe, as reported in Tab.~\ref{tab:channels}.
  Therefore, the number of different frequency channels for \gls{ul} is $C
  = 3$.
\item \textit{Duty cycle} -- The simulator considers the \gls{dc} limitations applied in the
  European region~\cite{regional}, which corresponds to setting $\delta_{SB1} = 99$ and $\delta_{SB2} = 9$ in the model.
\item \textit{Channel model} -- Differently from the model, simulated LoRaWAN
  nodes experience a log-distance propagation path loss,
  as for an open-air scenario. Thus, farther devices will suffer increased
  loss, and their performance will be penalized with respect to \glspl{ed} that
  are close to the \gls{gw}. Note that we do not include fast-fading components,
  which are supposed to be averaged out by the LoRa modulation, nor
  time-dependent variations in the channel, which remains constant throughout the
  entire simulation. Also, the channel is assumed to be symmetric, and
  \gls{dl} transmissions will suffer the same impairments as in the \gls{ul}.
  \item \textit{\gls{sf} distribution} -- \glspl{ed} are located around the
        single \gls{gw} in a circular area of radius 2500~m, which allows for
        communications with any SFs with negligible channel error probability
        (in the absence of interference). Instead, the positions of the nodes
        are randomly picked at each simulation run. \glspl{sf} are assigned
        uniformly (see Tab.~\ref{tab:resparams}, $p_{\rm equal}$). A different
        \gls{sf} distribution ($p_{\rm EXPLoRa}$) is considered in some
        scenarios, to evaluate the impact of this parameter on the different
        metrics.
  \item \textit{Interference and capture effect} -- To model interference, in
        the simulator we consider the collision matrix provided
        in~\cite{goursaud2015dedicated} and the overlapping time between
        packets, as described in~\cite{magrin2017performance}.\footnote{Note
        that, in the simulator, the capture event is determined also considering
        the partial overlapping of the colliding packets.} A packet survives
        interference from a signal modulated with the same \gls{sf} if its power
        is at least $CR_{dB} = 6$~dB higher than the colliding one. In order to
        provide a comparison with this scenario, in the analytical model we
        leverage the assumption of uniformly distributed \glspl{ed} around the
        \gls{gw} to compute the capture probabilities as
        in~\cite{bankov2017mathem}, which results in $\mathbb{W}^{GW} = 0.1796$,
        and $\mathbb{W}^{ED} = 0.5682 $. We remark that different distributions
        of \glspl{ed} around the \gls{gw} can be modeled by adapting this
        derivation.
\end{itemize}
%


Since the \gls{gw} implementation in the simulator attempts to emulate the
behavior of a real device, a \gls{ul} packet is successfully received when all the
following conditions are satisfied:
\begin{enumerate}
\item The packet finds an available demodulator;
\item The packet's reception is not interrupted by \gls{dl} transmissions;
\item Once the reception is finished, the packet was not corrupted by
  interference.
\end{enumerate}

To count packets at the \gls{phy} layer coherently with the simulator
implementation, the model's packet loss probabilities due to lack of
demodulators ($F_{NMD}$), \gls{gw} transmission ($F_{GWTX}$) and interference
($F_{INT}$) are plotted in the following section using, respectively, the
following expressions:
\begin{enumerate}
\item $F_{\rm NMD}= 1 - S^{demod}$;
\item $F_{\rm GWTX} = E_{i}\left[S^{demod} \cdot (1 - S_i^{TX})\right]$;
\item $F_{\rm INT} = E_{i}\left[S^{demod} \cdot S_i^{TX} \cdot (1 - S_i^{INT})\right]$;
\end{enumerate}
by exploiting~\eqref{eq:Sint},~\eqref{eq:Sitx}, and~\eqref{eq:Sdemod}, and where
$E_i\left[\cdot\right]$ indicates the expectation over the distribution of
\glspl{sf} and $S^{demod}$ the probability that, in the simulations, a packet
can lock on an available demodulator.


\section{Results}
\label{sec:results}

This section provides a comparison between the performance estimated with the
proposed model and by the ns-3 simulator. Results are presented for both
\gls{phy} and \gls{mac} layer, and the impact of the model's assumptions is
shown to be mostly negligible, or at least acceptable. Finally, some results
will showcase how the model can be used to gain insight on the behavior of the
LoRaWAN technology in a quick and effortless way, analyzing the effects of
various parameters on the performance of the network. In the plots of this
section the analytical results are represented by lines, while markers
correspond to simulation outcomes.

\begin{figure}[t]
	\centering
  \includegraphics[width=\figurescaling\linewidth]{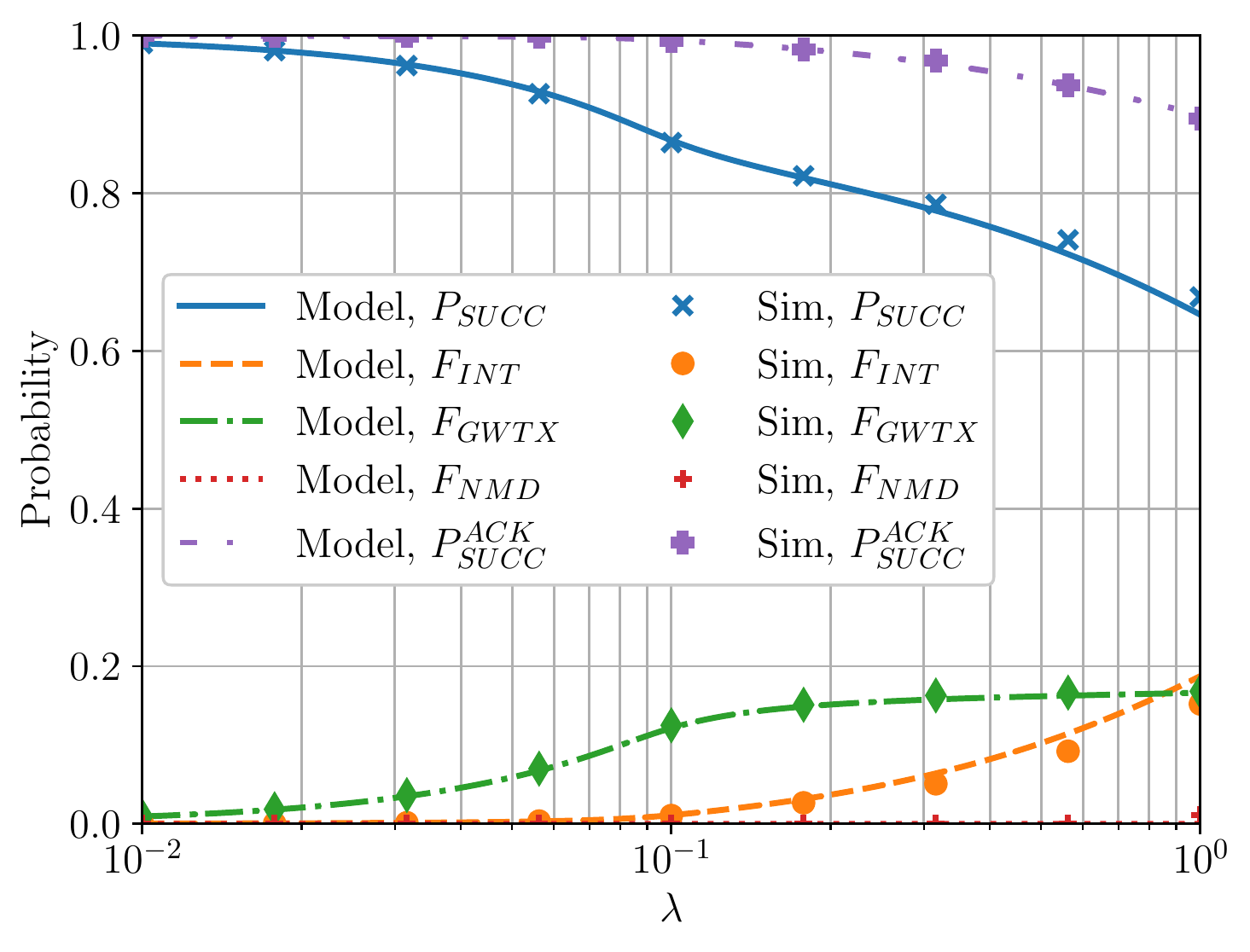}
	\caption{PHY-level performance with $m=8$, $\alpha=1$.}
	\label{fig:phy}
  \vspace{-1em}
\end{figure}
Fig.~\ref{fig:phy} shows the packet outcome probabilities at the \gls{phy}
layer in a network employing confirmed traffic. Although obtained with
different approaches, such probabilities are overall consistent, proving
the effectiveness of the model.

\begin{figure}[t]
	\centering
  \begin{subfigure}[t]{\figurescaling\linewidth}
    \centering
    \includegraphics[width=\linewidth]{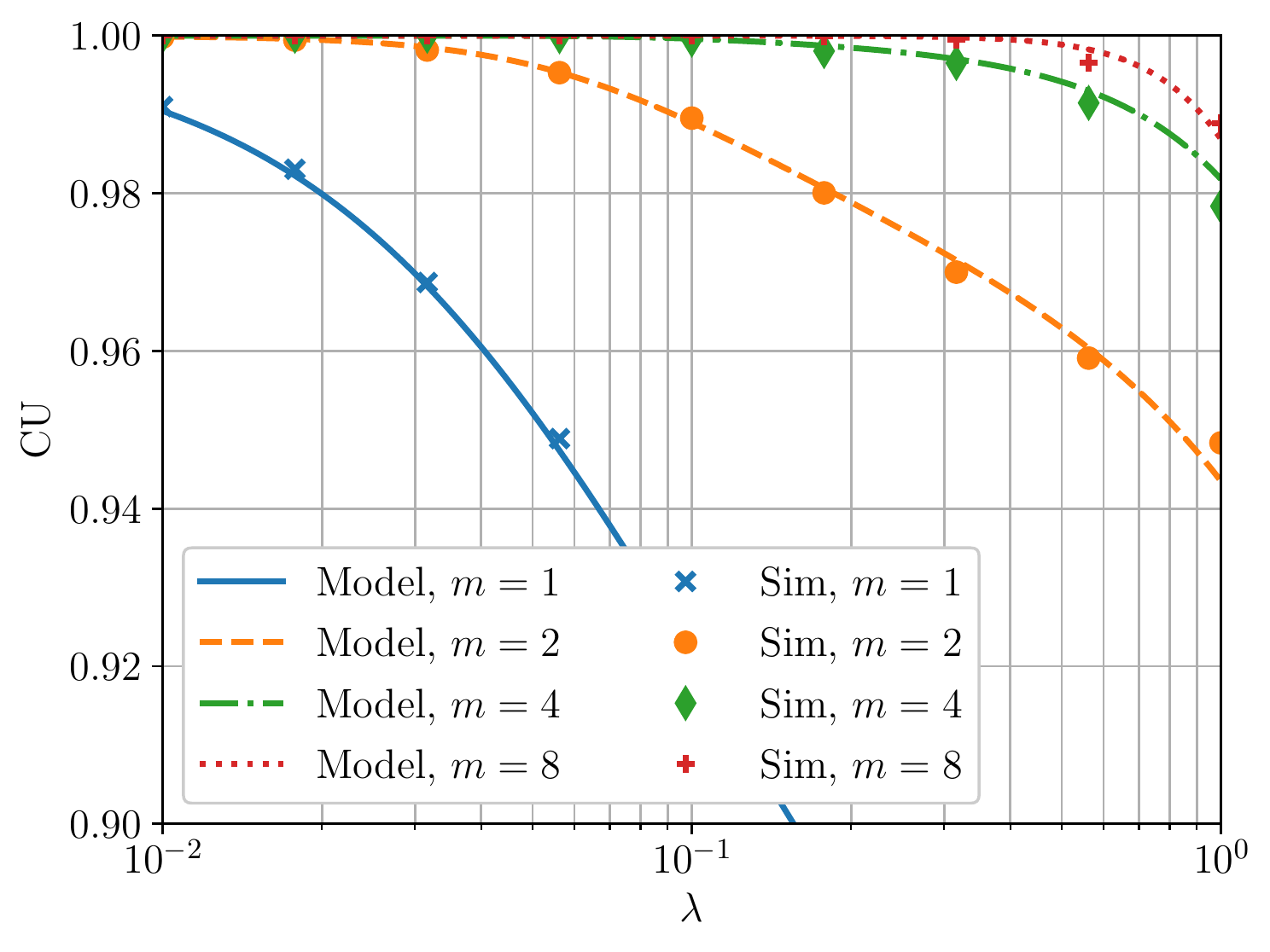}
    \vspace{-0.5cm}
    \caption{CU for different values of $m$, $\alpha=1$}
    \label{fig:cu}
  \end{subfigure}%
  \begin{subfigure}[t]{\figurescaling\linewidth}
    \centering
    \includegraphics[width=\linewidth]{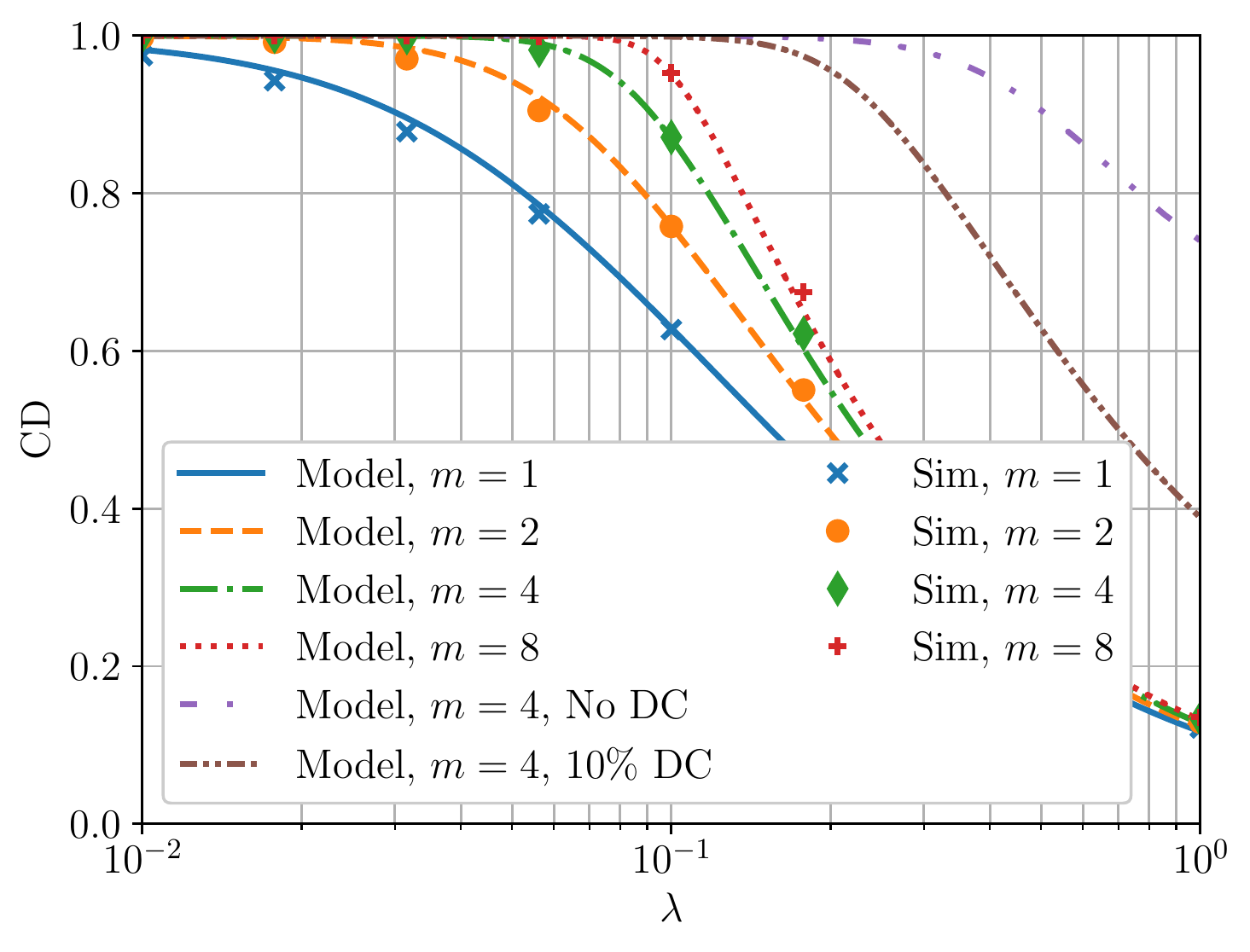}
    \vspace{-0.5cm}
    \caption{CD for different values of $m$, $\alpha=1$}
    \label{fig:cd}
  \end{subfigure}%
	\caption{Comparison of model and simulation results in terms of CU and
    CD.}
	\label{fig:cucd}
  \vspace{-1em}
\end{figure}

The good match between model and simulation is also reflected in
Fig.~\ref{fig:cucd}, which shows the \gls{cu} and \gls{cd} metrics for a network
in which all \glspl{ed} generate confirmed traffic ($\alpha=1$), and for
different values of $m$. Also in this case, the model results are quite close to
those given by the simulations. Fig.~\ref{fig:cu} shows that the number of
available transmissions helps the correct delivery of the message at the
\gls{mac} layer, providing performance above 0.9 also for relatively high
traffic levels, when an average of one packet per second is generated by the
network at the application layer. The \gls{cd} performance shown in
Fig.~\ref{fig:cd} exhibits a similar behavior, but reaches much lower values
mostly because the rate of \gls{dl} messages that the \gls{gw} can generate is
limited by the \gls{dc} restrictions. The fact that this loss in performance is
caused by the \gls{gw}'s \gls{dc} is confirmed by the lilac dash-dotted line in
Fig.~\ref{fig:cd}: to obtain these results, the \gls{dc} restrictions were
lifted by setting $\delta_{SB1} = \delta_{SB2} = 0$ in the model, producing
markedly better results when compared to the corresponding green curve, where
\gls{dc} is enabled. Another example of the model's flexibility in considering
also non-standard settings is given by the densely dash-dotted brown line, which
represents the \gls{cd} metric when $\delta_{SB1} = \delta_{SB2} = 9$, i.e.,
when transmissions in both sub-bands are subject to a \gls{dc} of 10\%. Although
being an ideal setting, this case shows that even a small increase in the
\gls{dc} allowance in SB1 can yield considerable performance gains.

\begin{figure}[t]
	\centering
  \includegraphics[width=\figurescaling\linewidth]{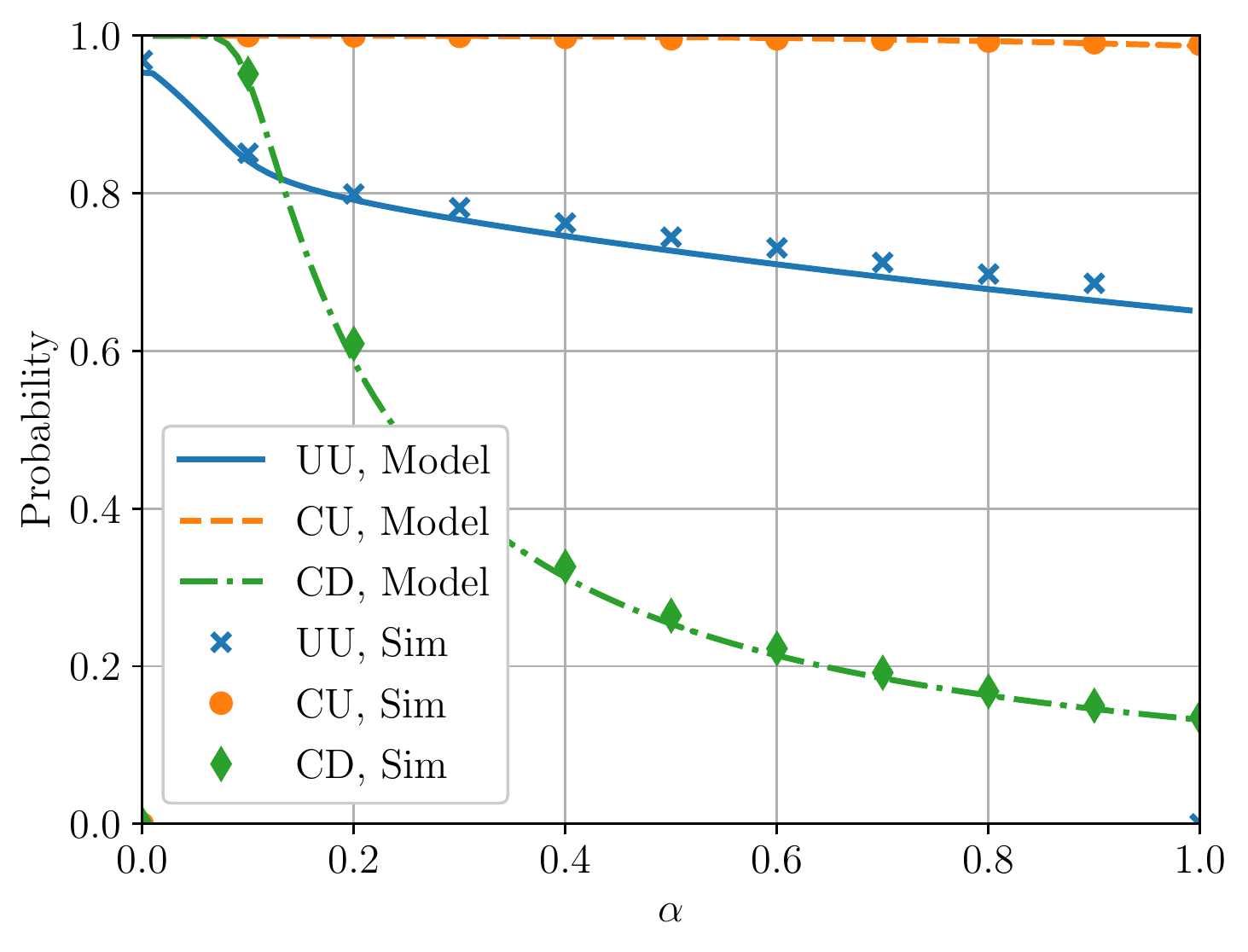}
	\caption{Performance when varying the fraction of confirmed traffic, with
    $\lambda=1, m=8, h=1$.}
	\label{fig:alpha}
  \vspace{-1em}
\end{figure}
Fig.~\ref{fig:alpha} compares simulation and theoretical results, in terms of
\gls{uu}, \gls{cu} and \gls{cd}, when different fractions of confirmed traffic
are employed in the network. For this comparison, we set the network application
layer packet arrival rate to $\lambda=1$ pck/s, the maximum number of
transmissions for confirmed traffic to $m=8$, and the number of repetitions for
unconfirmed traffic to $h=1$. As the fraction of \glspl{ed} employing confirmed
traffic increases, the data delivery performance decreases for all the
\glspl{ed}, in particular for nodes employing unconfirmed traffic which do not
have the chance of re-transmitting their packets. The match between the
simulator and the model is confirmed to be excellent for all values of $\alpha$.

\begin{figure}[t]
	\centering
  \begin{floatrow}
  \ffigbox{\includegraphics[width=0.95\linewidth]{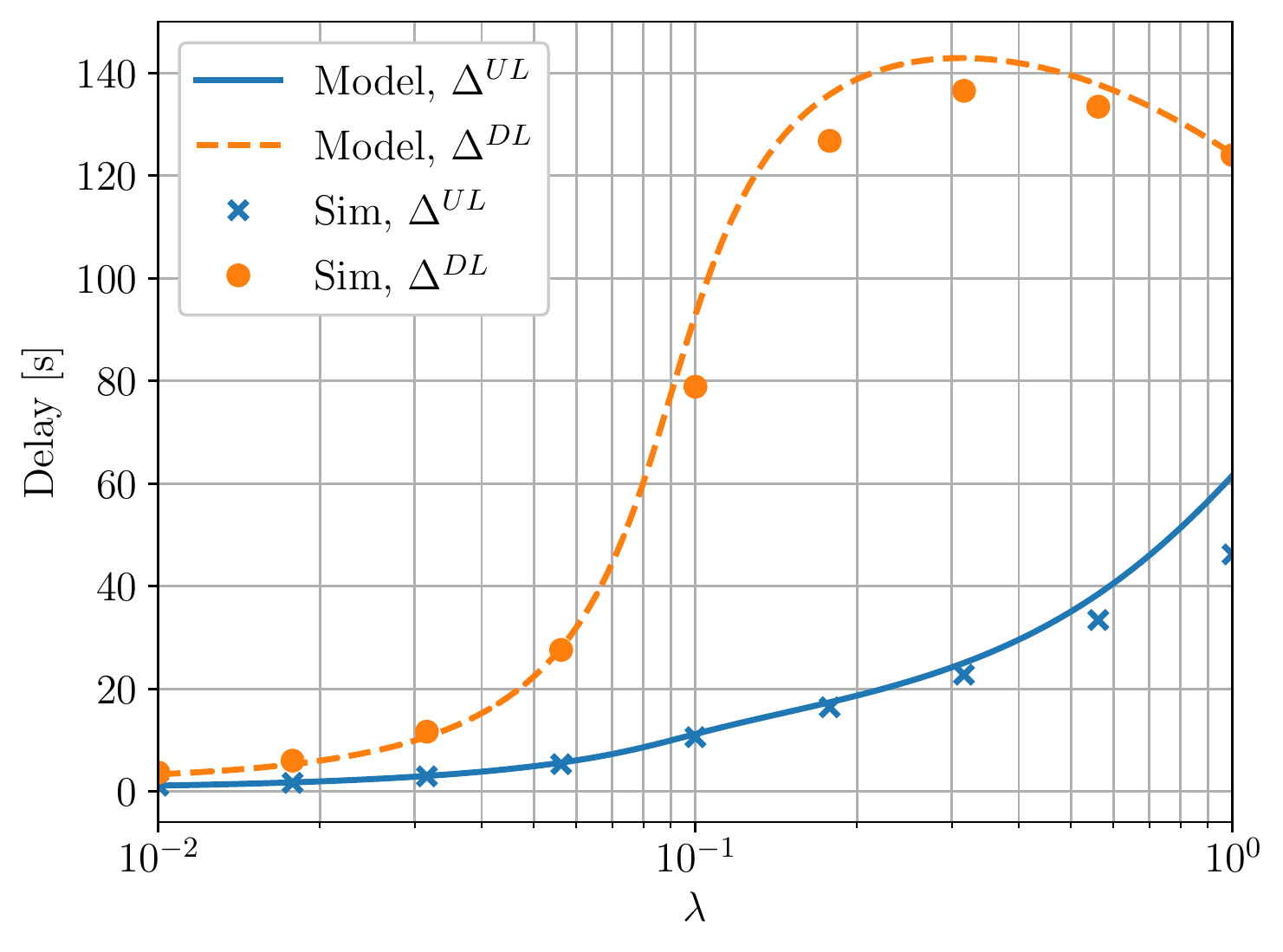}}{\caption{Delays for a confirmed traffic network, $m=8$.}\label{fig:delays}}
  \ffigbox{\includegraphics[width=0.95\linewidth]{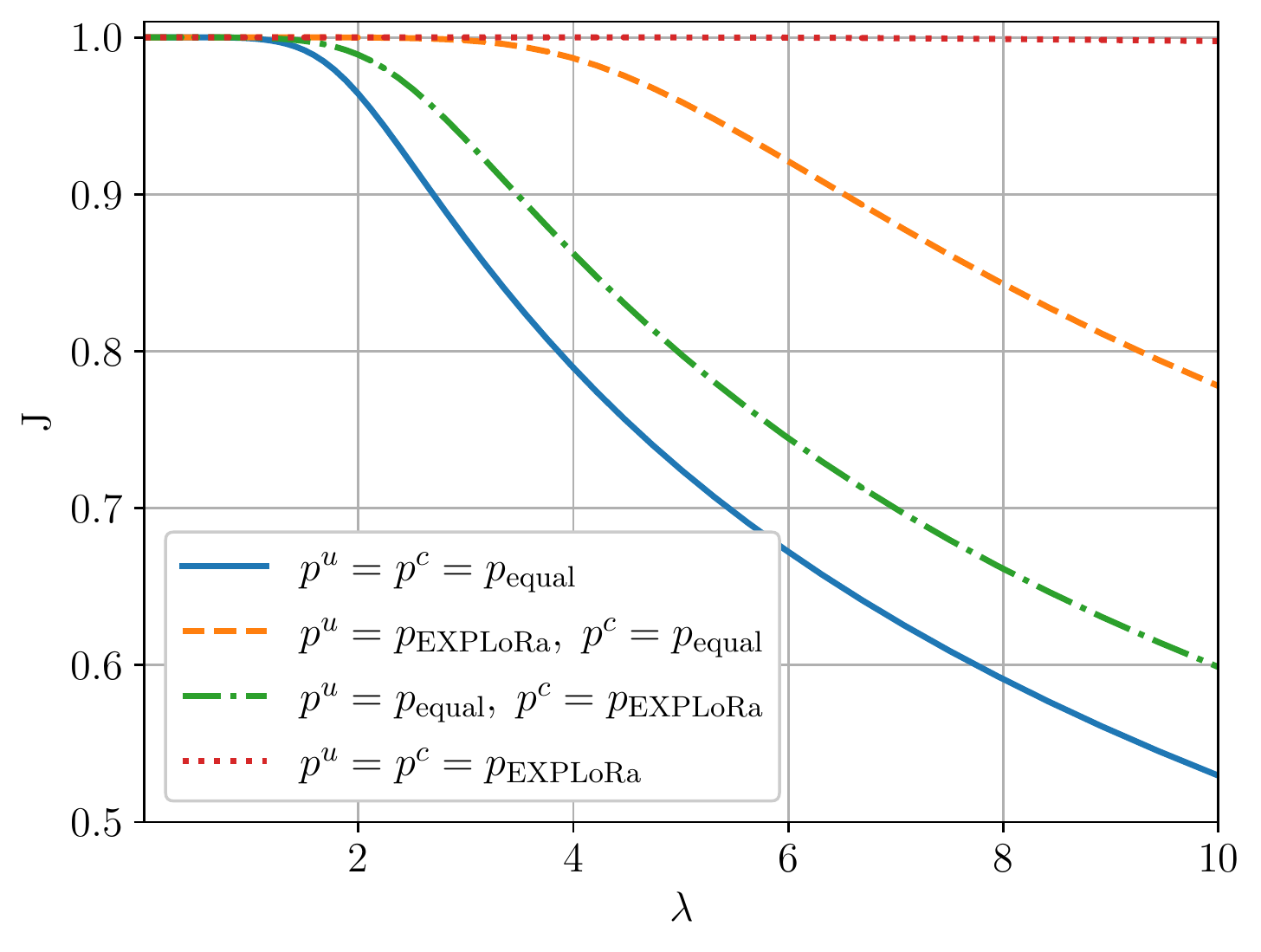}}{\caption{Fairness for different SF distributions when $m=8, h=8, \tau=1, \alpha=0.3$.}   \label{fig:fairness}}
  \end{floatrow}
\end{figure}
The final metric that we evaluate through both model and simulation is the
delay, as described in Sec.~\ref{sec:metrics}. Fig.~\ref{fig:delays} shows
how delays generally increase with the traffic load, since more
re-transmissions are needed to successfully deliver a packet. Note that for
high values of $\lambda$ the average \gls{ack} delay $\Delta^{\textrm{DL}}$
decreases: this is explained by the fact that devices employing higher
\glspl{sf}, (which may increase the average delay due to their longer
inter-packet transmission times) heavily suffer from interference and are
often dropped (unsuccessful packets are not considered in the delay
computation). Although not shown here, it is worth noting that the model
formulation makes it easy to extract per-\gls{sf} metrics that can help
troubleshoot the network configuration under study.

%
We now analyze how the fairness varies with the traffic load for different
configurations of $\alpha$, $p^u$ and $p^c$. We consider the \gls{sf}
distributions $p_{\rm equal}$ and $p_{\rm EXPLoRa}$ as defined in
Tab.~\ref{tab:config}. The $p_{\rm EXPLoRa}$ distribution, first presented
in~\cite{cuomo2017explora}, aims at equalizing the aggregate time on air of each
group of devices employing the same \gls{sf} to minimize the collision
probability. In Fig.~\ref{fig:fairness} we can observe that, when the \glspl{sf}
are uniformly allocated independently of the traffic type (i.e.,
$p^u = p^c = p_{\rm equal}$), the fairness decreases for an increasing traffic
intensity. Indeed, as the traffic grows, nodes employing lower \glspl{sf} will
suffer less from interference because of the shorter transmission times. The
fairness grows when $\alpha=0.3$ and $p^c = p_{\rm EXPLoRa}$, since with this
configuration 30\% of the generated packets will use lower \glspl{sf} with
higher probability, diminishing the channel and \gls{gw} occupancy. However,
since the traffic load is high and the fairness is measured on the uplink
performance (\gls{uu} and \gls{cu}), the beneficial effect of allocating
\glspl{sf} according to the $p_{\rm EXPLoRa}$ distribution are more evident when
it is used for most of the devices, i.e., the 70\% of nodes employing
unconfirmed traffic. Finally, the maximum fairness is achieved when the
\glspl{sf} are allocated using $p_{\rm EXPLoRa}$ both for $p^u$ and $p^c$
(dotted line in
Fig.~\ref{fig:fairness}). 
Note that, when $\lambda \leq 1$, the load in the network is low enough to have
$J=1$ for every $p^u, p^c$, since the collision probability is low and the \gls{gw} is not busy with
\gls{ack} transmissions.

\begin{figure}[t]
  \centering
  \includegraphics[width=\figurescaling\linewidth]{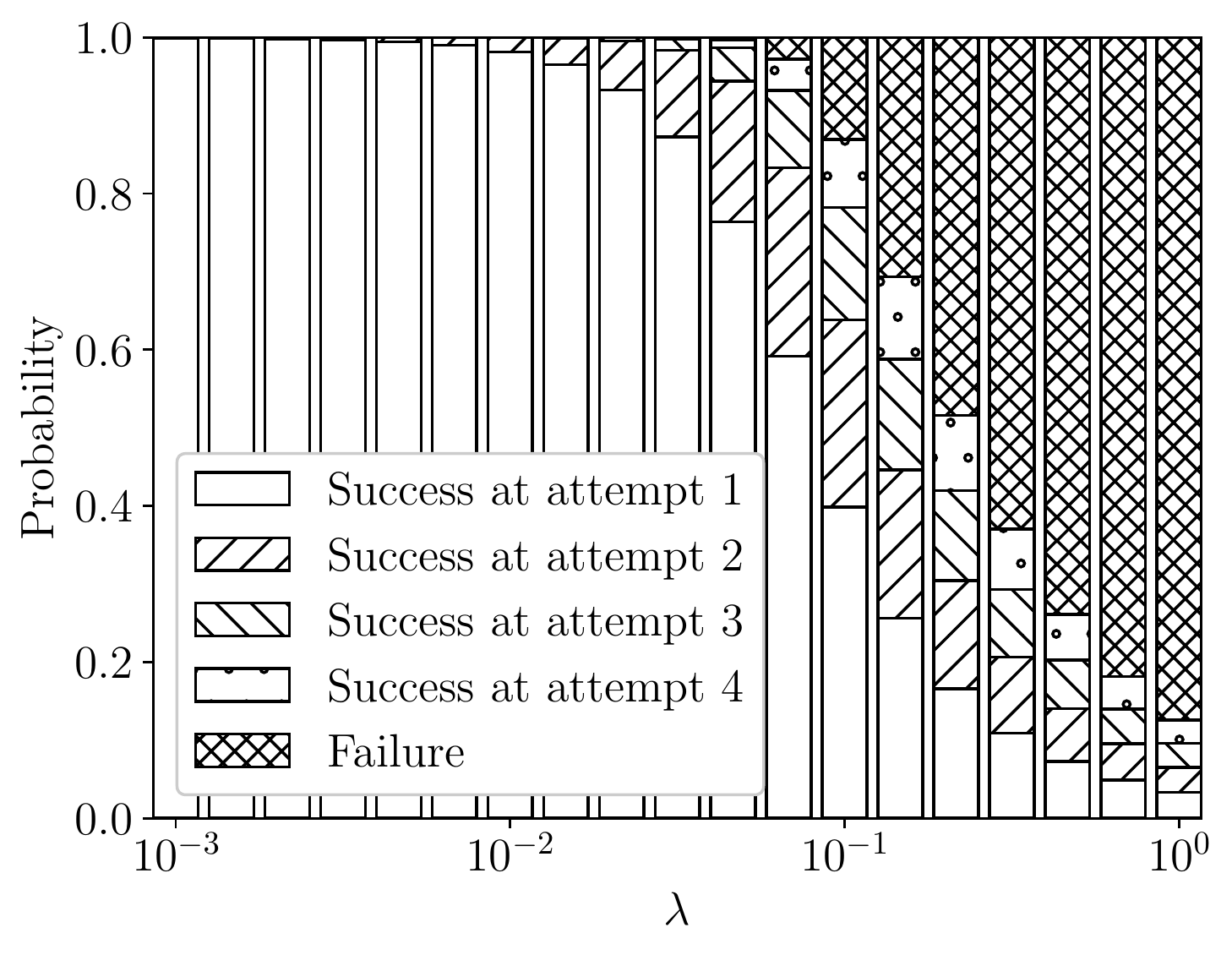}
  \caption{Distribution of re-transmissions, $m=4$, $\alpha=1$.}
  \label{fig:retxdistribution}
  \vspace{-1em}
\end{figure}

An example of insight that the analytical model can offer is presented in
Fig.~\ref{fig:retxdistribution}, which shows the fraction of traffic that
achieves success after a certain number of re-transmission attempt for different
traffic loads, derived from $P^{DL}_{i,j}$. This data, for instance, can be used
to estimate the power consumption at the nodes: for low traffic loads the vast
majority of \gls{mac} layer packet transmissions succed with just one \gls{phy}
layer transmission attempt. As the traffic load increases, the fraction of
devices needing multiple re-transmissions to correctly receive an \gls{ack}
correspondingly increases. After a certain point, packet reception fails with
such a high rate that most \glspl{ed} need to employ the maximum number of
transmissions and, despite the high energy expenditure, still fail to receive an
\gls{ack} from the \gls{gw}.

\begin{table}
  \footnotesize
  \centering
  \caption{Configurations employed in Fig.~\ref{fig:improvements}}
  \label{tab:config}
  \begin{tabular}{lccccc}
    \toprule
    Configuration & $\tau_1$ & $\tau_2$ & $m$ & $h$ & $p^u = p^c$ \\
    \midrule
    C1 & 1 & 1 & 1 & 1 & $p_{\rm equal}$   \\
    C2 & 0 & 1 & 1 & 4 & $p_{\rm EXPLoRa}$ \\
    C3 & 0 & 1 & 4 & 4 & $p_{\rm EXPLoRa}$ \\
    \bottomrule
  \end{tabular}
\end{table}
\begin{figure}[t]
	\centering
  \includegraphics[width=\figurescaling\linewidth]{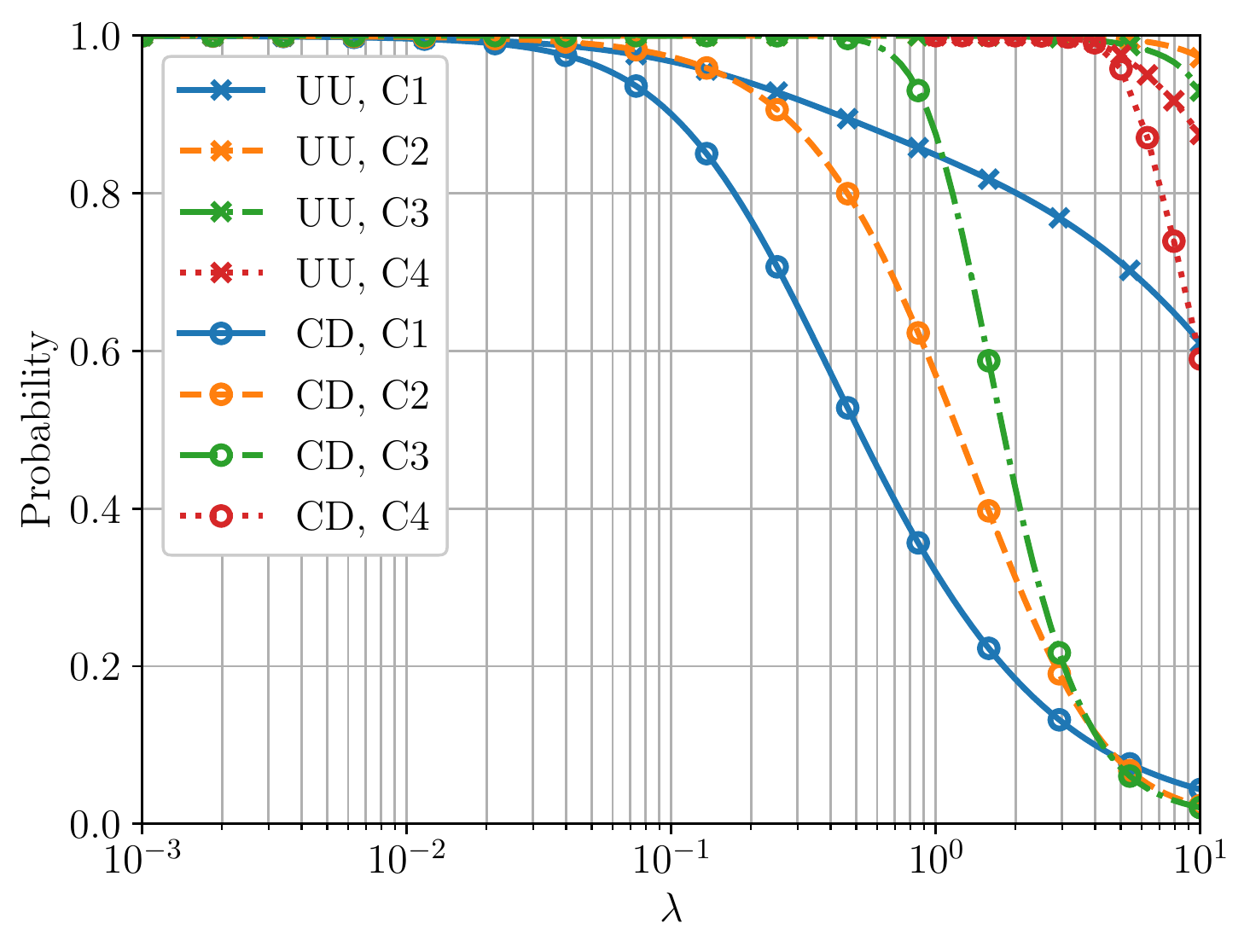}
	\caption{UU and CD performance for different network configurations,
    $\alpha=0.3$.}
	\label{fig:improvements}
  \vspace{-1em}
\end{figure}
Finally, we show how the model can be applied to investigate the impact of
different network parameters on the performance. In the example of
Fig.~\ref{fig:improvements}, 30\% of the \glspl{ed} employ confirmed traffic,
and we show results obtained with the proposed mathematical model. The parameter
configurations are summarized in Table~\ref{tab:config}. Configuration C1
provides a baseline: priority is given to \gls{dl} transmission in both windows,
devices employ a single transmission attempt for both confirmed and unconfirmed
traffic, and \glspl{sf} are uniformly distributed. In this case the curves have
a shape similar to those shown in Fig.~\ref{fig:cucd} for $m=1$, but, since
fewer devices require \glspl{ack}, the \gls{gw} is able to receive more packets
and profitably send replies, leading to better performance. To improve \gls{uu}
a second configuration (C2) considers the prioritization of \gls{ack}
transmissions in \gls{rx2}, where their reception suffers less interference.
Moreover, unconfirmed packets are sent multiple times and we use $p^u = p^c =
p_{\rm EXPLoRa}$. This configuration provides a considerable
improvement with respect to the \gls{uu} metric, and some gains are also
achieved in the \gls{cu} performance. To improve also the results for confirmed
traffic, a further step (configuration C3) is to set $m=4$. This provides a
significant improvement of \gls{cu}, at the cost of a (minimal) decrease in
\gls{uu} performance. As a final step, we fully leverage the analytical model to
identify the optimal parameter configuration (i.e., $m$, $h$, $p_{u}$ and
$p_{c}$) for each plotted traffic load, with the objective of maximizing the
average of \gls{uu} and \gls{cu}. The red curves of this setting (C4) show how
this optimization process enabled by the model can significantly improve the
global performance of the network, significantly improving the \gls{cd}
performance at the price of a very small reduction in packet success rate for
unconfirmed devices.

The optimization problem that is solved to obtain configuration C4 is defined
as:
\begin{equation}
\begin{aligned}
\max_{p_{u}, p_{c}} \quad & \textrm{UU} + \textrm{CD} \\
\textrm{s.t.} \quad     & 0 \le p_{i}^{u} \le 1 \\
                        & 0 \le p_{i}^{c} \le 1 \\
                        & \sum_{i} p_{i}^{u} = 1 \\
                        & \sum_{i} p_{i}^{c} = 1 \\
\end{aligned}
\label{eq:optimization}
\end{equation}
where we explore the entire space defined by $m$, $h$ and $\lambda$, by solving~\eqref{eq:optimization}
to find the best $p_{u}$ and $p_{c}$, and finally pick the best solution for
each $\lambda$. The search is performed using the trust region method as
implemented by the \texttt{scipy} library, and we always set $p_{i}^{u} =
p_{i}^{c} = 1/6$ as the initial parameter value for the algorithm.

\begin{figure}[t]
	\centering
  \includegraphics[width=0.6\linewidth]{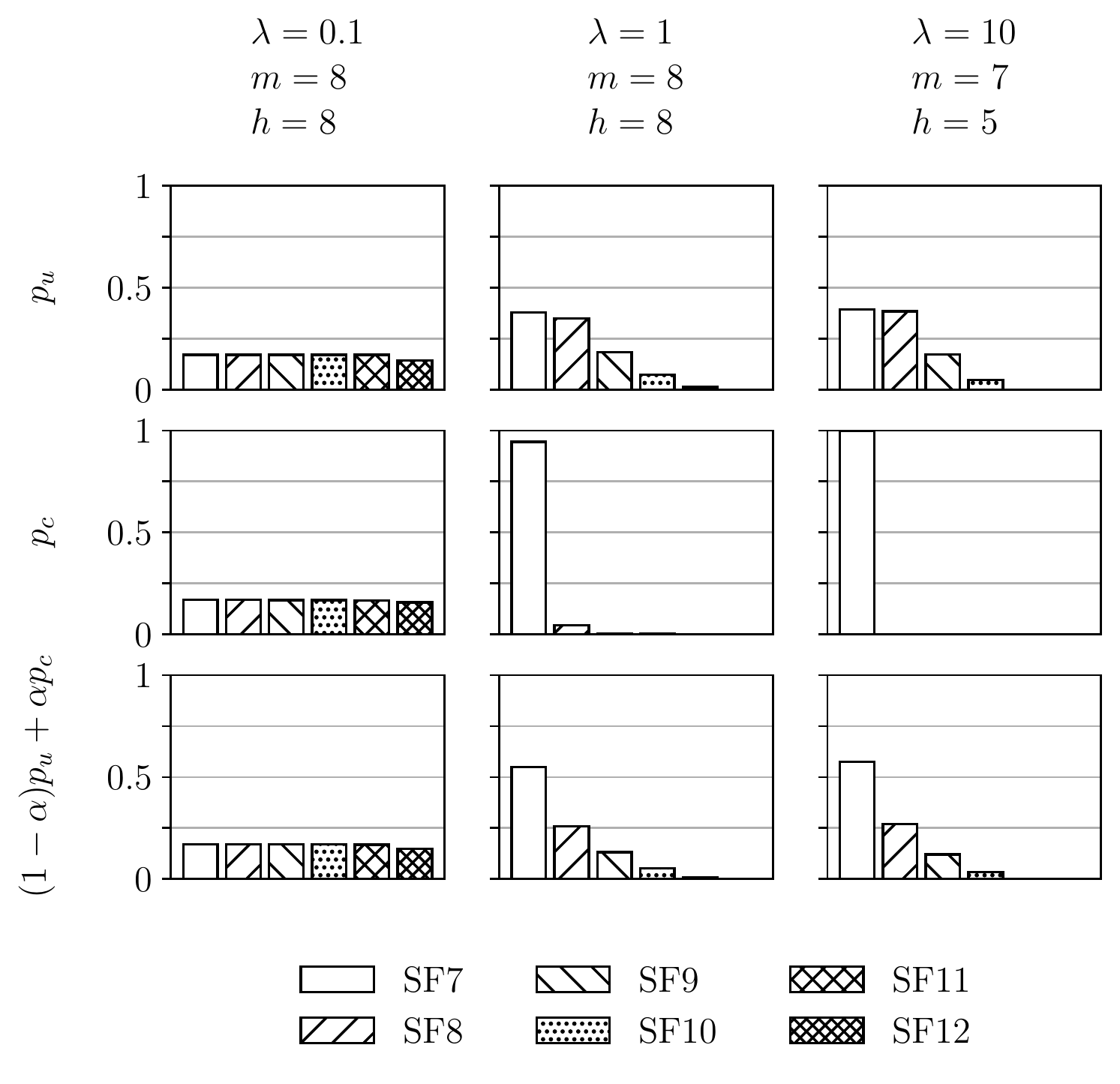}
	\caption{Optimal values of $p_{u}$, $p_{c}$, $m$ and $h$ as computed through model-driven optimization, for various values of $\lambda$.}
	\label{fig:optimal-parameters}
  \vspace{-1em}
\end{figure}
Figure~\ref{fig:optimal-parameters} displays the parameters of configuration C4
for some representative values of $\lambda$, showing $p_{u}$ in the first row,
$p_{c}$ in the second row, and a combination of the two weighed on $\alpha$ on
the third row. For a low value of generated traffic ($\lambda = 0.1$, first
column), we see that the optimization stops almost immediately, yielding a
distribution that is very similar to the initial value of $p_{u}$ and $p_{c}$.
In this case, as can also be seen in Figure~\ref{fig:improvements}, since the
traffic load is low the performance is indeed very good for high values of $m$
and $h$, and needs little optimization of the \gls{sf} distributions. For
$\lambda = 1$, instead, the optimization process yields a more distinctive value
of $p_{c}$, setting almost all devices to use \gls{sf}7. This is motivated by
the fact that, \gls{rx1} is set to employ the same \gls{sf} used in the
\gls{ul}. Therefore, having most of the confirmed devices employ an \gls{sf} as
low as possible is advantageous, since it guarantees faster \gls{ack}
transmissions in the \gls{dl} and, as a consequence, shorter silent times
imposed by the \gls{dc}, and a larger set of devices can thus be served. Devices
employing unconfirmed traffic, instead, are set to use a variety of \gls{sf}
values. Notably, the selected values are such that the aggregated distribution
considering both unconfirmed and confirmed traffic (visible in the third row)
takes a shape that is very similar to that of $p_{\rm EXPLoRa}$. This behavior
is even more marked when $\lambda = 10$, with the notable difference that higher
\gls{sf} values are not used in the optimized network: this is because of the
limited number of demodulators at the \gls{gw} (a factor which is accounted for
in our model). Indeed, although using all \gls{sf} values would bring an
additional gain, a packet with high \gls{sf} value occupies a demodulator for
quite a long time, increasing the probability that other incoming packets are
dropped because of unavailability of reception chains at the \gls{gw}. Finally,
we note that $m$ and $h$ are consistently set to their maximum values (8 here)
up to $\lambda = 1$. After this value, instead, it pays off to reduce the number
of repetitions employed by both unconfirmed and confirmed \glspl{ed}.

Although this analysis showcases the potential of the mathematical model to
identify the optimal settings, an evaluation of the trade-offs associated to
parameter configurations and their effect on other metrics of interest, such as
delays and energy consumption, needs a deeper investigation, which we leave for
future work.

%

\section{Conclusion}
\label{sec:conclusion}

In this work, we presented a model for the performance evaluation of a
LoRaWAN network in the presence of both confirmed and unconfirmed traffic,
taking into account the influence of different settings of multiple network
configuration parameters.

The model is able to capture both the \gls{phy} layer and \gls{mac} layer
performance, and describes the multiple events that affect both \gls{ul} packet
reception and \gls{dl} transmission: interference, capture effect, availability
of demodulator, \gls{dc} constraints, ongoing transmissions and receptions.
We validated the model results with ns-3 simulations, showing the consistency
among the two sets of results. Finally, we presented some examples of how the
model can be employed to analyze the effects of possible changes to the standard
parameter settings, and to identify optimal configurations with minimum effort.

Several extensions of this work are possible. A first improvement to the model
is the inclusion of multi-\gls{gw} scenarios, where \gls{ul} packets are
potentially received by several \glspl{gw}, and the network \gls{dl} capacity is
increased. A second aspect of interest is to leverage the proposed model to
better investigate trade-offs among different network parameters in various
scenarios, or when specific performance requirements are provided. A third
possible improvement would involve characterizing the capture effect for
non-uniform spatial distribution of the devices. Finally, a fourth direction is
to employ the proposed model to identify optimal network settings when different
metrics of interest are used as optimization functions, as we showed in the
results section with some simple cases. We point out that the target of the
model was to explore the capabilities of LoRaWAN networks, thus, in this work,
we neglected some features of LoRa, such as the interference between overlapping
packets modulated with different \glspl{sf}. The model can be extended by
including this, as well as other specific features of the LoRa technology. Such
extensions are left for future work.

We remark that all figures contained in this paper, covering both
model evaluations and simulation results, can be easily reproduced using
the tool available at~\cite{publishedmodelcode}.

\section*{Acknowledgment}

Part of this work was supported by MIUR (Italian Ministry for Education and
Research) under the initiative "Departments of Excellence" (Law 232/2016).

\bibliographystyle{IEEEtran} \bibliography{refs}

\end{document}